\documentclass[a4paper,aps,prd,twocolumn,10pt,superscriptaddress,nofootinbib,showpacs]{revtex4-1}
\usepackage{graphicx,amsmath,amsfonts,amssymb,revsymb,dcolumn,epsfig,bm,xspace}
\usepackage[hyperfootnotes=false]{hyperref}
\usepackage[latin1]{inputenc}
\usepackage{color}

\newcommand{\dd}{\mathrm{d}}
\newcommand{\lie}{\pounds}
\newcommand{\del}{\partial}
\newcommand{\dirac}[2]{\delta^{#1}(#2)}
\newcommand{\Poisson}[2]{\left\{#1,\;#2\right\}}
\newcommand{\indspace}{\quad}
\DeclareFontFamily{U}{mathx}{\hyphenchar\font45}
\DeclareFontShape{U}{mathx}{m}{n}{<-> mathx10}{}
\DeclareSymbolFont{mathx}{U}{mathx}{m}{n}
\DeclareMathAccent{\widebar}{0}{mathx}{"73}
\newcommand{\backdec}[1]{\bar{#1}}
\newcommand{\wbackdec}[1]{\widebar{#1}}
\newcommand{\g}{g}
\newcommand{\dg}{{\delta\g}}
\newcommand{\bg}{{\backdec{g}}}
\newcommand{\n}{v}
\newcommand{\bn}{{\backdec{\n}}}
\newcommand{\dn}{{\delta\n}}
\newcommand{\sn}{\mathsf{v}}
\newcommand{\h}{\gamma}
\newcommand{\bh}{\backdec{\h}}
\newcommand{\dhh}{{\delta\h}}
\newcommand{\hp}[1]{\h\left[#1\right]}
\newcommand{\bhp}[1]{\bh\left[#1\right]}
\newcommand{\cd}{\nabla}
\newcommand{\bcd}{\wbackdec{\nabla}}
\newcommand{\scd}{D}
\newcommand{\bscd}{\wbackdec{\scd}}
\newcommand{\scp}{\parallel}
\newcommand{\blap}{\wbackdec{\scd}^2}
\newcommand{\blapK}{\blap_{\K}}
\newcommand{\bilap}{\wbackdec{\scd}^{-2}}
\newcommand{\bilapK}{\bilap_{\K}}
\newcommand{\R}{R}
\newcommand{\dR}{{\delta\R}}
\newcommand{\bR}{\wbackdec{\R}}
\newcommand{\G}{G}
\newcommand{\bG}{\wbackdec{\G}}
\newcommand{\kp}{\kappa}

\newcommand{\ddg}{\xi}

\newcommand{\LL}{{\mathcal F}}
\newcommand{\LLa}{\LL_{a}{}}
\newcommand{\LLb}{\LL_{b}{}}
\newcommand{\SLL}{\Upsilon}
\newcommand{\SLLa}{\SLL_{a}{}}
\newcommand{\SLLb}{\SLL_{b}{}}
\newcommand{\A}{\phi}
\newcommand{\B}{B}
\newcommand{\C}{C}
\newcommand{\CS}{\psi}
\newcommand{\CSI}{\Psi}
\newcommand{\CSvms}{\mathcal{U}}
\newcommand{\MS}{\zeta}
\newcommand{\CST}{\CS^{\mathsf{t}}}
\newcommand{\CTD}{W}
\newcommand{\BS}{\mathcal{\B}}
\newcommand{\BV}{\mathtt{\B}}
\newcommand{\CSD}{\mathcal{E}}
\newcommand{\CV}{\mathtt{F}}

\newcommand{\EC}{\mathcal{K}}
\newcommand{\bEC}{\wbackdec{\EC}}
\newcommand{\dEC}{{\delta\EC}}
\newcommand{\EX}{\Theta}
\newcommand{\dEX}{{\delta\EX}}
\newcommand{\dEXi}{\Xi}
\newcommand{\bEX}{\wbackdec{\EX}}
\newcommand{\SH}{\sigma}
\newcommand{\dSH}{\delta\sigma}
\newcommand{\dSHs}{\dSH^{\text{s}}{}}
\newcommand{\dSHv}{\dSH^{\text{v}}{}}
\newcommand{\bSH}{\wbackdec{\SH}}
\newcommand{\SR}{\mathcal{R}}
\newcommand{\dSR}{{\delta\SR}}
\newcommand{\bSR}{\wbackdec{\SR}}
\newcommand{\K}{K}
\newcommand{\bK}{\wbackdec{\K}}
\newcommand{\ac}{a}
\newcommand{\bac}{\backdec{\ac}}
\newcommand{\dac}{{\delta\ac}}
\newcommand{\gH}{\mathcal{H}}
\newcommand{\T}{T}

\newcommand{\bT}{\wbackdec{T}}
\newcommand{\dT}{\delta\T}
\newcommand{\mf}{\varphi}
\newcommand{\dmf}{\delta\mf}
\newcommand{\nm}{u}
\newcommand{\vm}{U}
\newcommand{\vms}{\mathcal{V}}

\newcommand{\ST}{\Sigma}
\newcommand{\dST}{{\delta\ST}}
\newcommand{\bST}{\wbackdec{\ST}}
\newcommand{\ent}{\vartheta}
\newcommand{\denta}{{\delta\ent}}
\newcommand{\bent}{\backdec{\ent}}
\newcommand{\en}{s}
\newcommand{\Pben}{\mathcal{S}}
\newcommand{\dent}{\delta\en{}}
\newcommand{\ben}{\backdec{\en}}
\newcommand{\den}{n}
\newcommand{\dden}{{\delta\den}}
\newcommand{\bden}{\backdec{\den}}
\newcommand{\p}{p}
\newcommand{\dpp}{\delta\p}
\newcommand{\temp}{\tau}
\newcommand{\btemp}{\backdec{\temp}}
\newcommand{\ed}{\rho}
\newcommand{\ved}{\varrho}
\newcommand{\cs}{c_s}
\newcommand{\bcs}{\backdec{c}_s}
\newcommand{\tal}{\iota}
\newcommand{\btal}{\backdec{\tal}}
\newcommand{\qual}{\varpi}
\newcommand{\bqual}{\backdec{\qual}}
\newcommand{\ded}{\delta\ed}
\newcommand{\dved}{\delta\ved}
\newcommand{\dq}{\delta{}q}
\newcommand{\fp}{\varphi}
\newcommand{\fpa}{{\fp_1{}}}
\newcommand{\fpb}{{\fp_2{}}}
\newcommand{\fpc}{{\fp_3{}}}
\newcommand{\fpd}{{\fp_4{}}}
\newcommand{\dfp}{\delta\fp}
\newcommand{\dfpa}{{\delta\fpa}}
\newcommand{\dfpb}{{\delta\fpb}}
\newcommand{\dfpc}{{\delta\fpc}}
\newcommand{\dfpd}{{\delta\fpd}}
\newcommand{\bfpa}{{\backdec{\fpa}}}
\newcommand{\bfpb}{{\backdec{\fpb}}}
\newcommand{\bfpc}{{\backdec{\fpc}}}
\newcommand{\bfpd}{{\backdec{\fpd}}}
\newcommand{\tdiff}[3]{\left.\frac{\del{}#1}{\del{}#2}\right\vert_{#3}}
\newcommand{\bp}{\backdec{p}}
\newcommand{\bed}{\backdec{\ed}}

\newcommand{\bq}{\backdec{q}}
\newcommand{\zz}{\mathsf{z}}
\newcommand{\uu}{\mathsf{u}}
\newcommand{\ms}{\mathsf{v}}
\newcommand{\LA}{\mathcal{L}}

\newcommand{\AC}{S}
\newcommand{\bLA}{\wbackdec{\LA}}
\newcommand{\bAC}{\wbackdec{\AC}}

\newcommand{\dLA}{{\delta\LA}}
\newcommand{\dAC}{{\delta\AC}}
\newcommand{\sur}{\text{sur}}
\newcommand{\pot}{\text{p}}
\newcommand{\kin}{\text{k}}
\newcommand{\mat}{\text{m}}
\newcommand{\tbg}{\text{bg}}
\newcommand{\HA}{\mathcal{H}}
\newcommand{\dHA}{{\delta\HA}}
\newcommand{\pCSD}{\Pi_{\CSD}}
\newcommand{\pCST}{\Pi_{\CST}}
\newcommand{\pvms}{\Pi_{\vms}}
\newcommand{\pdent}{\Pi_{\dent}}
\newcommand{\pdfpc}{\Pi_{\dfpc}}
\newcommand{\pMS}{\Pi_\MS}
\newcommand{\pa}{\Pi_a}
\newcommand{\pbfpa}{\Pi_{\bfpa}}
\newcommand{\pbfpc}{\Pi_{\bfpc}}
\newcommand{\pben}{\Pi_{\ben}}
\newcommand{\BGE}{E}
\newcommand{\cV}{\widetilde{V}}
\newcommand{\cK}{\widetilde{\K}}

\begin{document}
\title{Quantum Cosmological Perturbations of Generic Fluids in Quantum Universes}

\author{S.~D.~P.~Vitenti} \email{vitenti@cbpf.br}
\author{F.~T.~Falciano} \email{ftovar@cbpf.br}
\author{N.~Pinto-Neto} \email{nelson.pinto@pq.cnpq.br}

\affiliation{Centro Brasileiro de Pesquisas F\'{\i}sicas -- CBPF,
Rua Dr.\ Xavier Sigaud 150, Urca, \\
CEP 22290-180, Rio de Janeiro -- RJ, Brasil}

\date{\today}

\begin{abstract}

In previous works, it was shown that the Lagrangians and Hamiltonians of cosmological linear scalar, vector and tensor perturbations of homogeneous and isotropic space-times with flat spatial sections containing a perfect fluid can be put in a simple form through the implementation of canonical transformations and redefinitions of the lapse function, without ever using the background classical equations of motion. In this paper, we generalize this result to general fluids, which includes entropy perturbations, and to arbitrary spacelike hyper-surfaces through a new method together with the Faddeev-Jackiw procedure for the constraint reduction. A simple second order Hamiltonian involving the Mukhanov-Sasaki variable is obtained, again without ever using the background equations of motion.

\end{abstract}


\maketitle

\section{Introduction}

In cosmology, when the curvature scale approaches the Planck length, one cannot avoid to consider quantum corrections already on the background geometry. In this regime, the usual semi-classical treatment of cosmological perturbations, where only the perturbations are quantized and the background is held classical, is no longer valid. Even though quantizing simultaneously the homogeneous background and their linear perturbations is still far from the full theory of quantum gravity, one can consider the inclusion of quantum effects in the dynamics of the background homogeneous model as an important improvement to the usual semi-classical approach~\cite{Halliwell1985}. Note, however, that this program prevent us from using the classical background equations, as it is usually done, to turn the full second order action into a simple treatable system. Notwithstanding, it has already been shown that it is possible to simplify this action through canonical transformation techniques for a barotropic perfect fluid and scalar fields without potential in a flat spatial section Friedmann model~\cite{Peter2005,Pinho2007,Falciano2009}.

The obtained Hamiltonian, with its zeroth and second order terms, was used to perform a canonical quantization yielding a Wheeler-DeWitt equation where cosmological perturbations of quantum mechanical origin evolve in a non-singular homogeneous and isotropic background in which quantum effects replace the usual classical singularity by a bounce near the Planck scale. The physical properties of these cosmological models were analyzed in many papers \cite{Peter2006,Peter2007,Peter2008}, and they proved to be complementary or even competitive with usual inflationary models as long as they are capable to lead to almost scale invariant spectra of long-wavelength cosmological perturbations.

The aim of this paper is to improve the previous formalism and to extend the known results to a generic thermodynamic fluid in a Friedmann background with arbitrary spacelike hyper-surfaces. In order to carry out this work, still without ever using the background classical equation, it proved to be simpler to work in the Lagrangian formalism, using a set of variable transformations along with Faddeev-Jackiw~\cite{Faddeev1988,Jackiw1993} reduction method rather than the Dirac formalism. One of the steps of the procedure was to isolate the perturbation terms in the action which are multiplied by the background classical equations of motion and eliminate them through some suitable field redefinitions, from where the Mukhanov-Sasaki variable naturally appears. The resulting action and Hamiltonian up to second order then become very simple and suitable for canonical quantization.

Besides the motivation related to the quantization procedure, our choice of variables used to write down the second order Lagrangian simplifies significantly the calculations involved. This simplification allows us to obtain all expressions without choosing a gauge. This is an important advantage since we have shown in Ref.~\cite{Vitenti2012} that the choice of a gauge implies an additional assumption that the perturbations remain small in this gauge.

The initial works which obtained the second order actions and Hamiltonians for perturbations in a background geometry~\cite{Mukhanov1981,Starobinsky1982,Hawking1982,Guth1982,Halliwell1985} were extended afterwards to the case of different matter contents~\cite{Langlois1994,Anderegg1994,Maldacena2003,Langlois2008,Deffayet2010,Kobayashi2010,Felice2010}. However, the majority of the papers cited above used the background equations of motion in order to simplify their final actions and Hamiltonians, and in the case they do not use them, their actions were either too complicated and intractable, or they were not general enough to include the cases of general fluids and/or curved spacelike hypersurfaces. The present paper fill this gap: we obtain a simple Hamiltonian for linear cosmological perturbations involving general fluids and curved spacelike hypersurfaces, without ever using the background equations of motion. It is worth emphasizing that our procedure differs significantly from the cited above. In our calculations we first expand the Lagrangian up to second order without choosing any spacetime foliation or background metric by writing the second order part quadratic in the variable $$\LL_{\alpha\beta}{}^\gamma \equiv -\frac{\g^{\gamma\sigma}}{2}\left(\g_{\sigma\beta;\alpha} + \g_{\sigma\alpha;\beta} - \g_{\alpha\beta;\sigma}\right),$$ where the semi-colon represents the covariant derivative with respect to the background metric. Afterwards, we express this tensor in terms of kinetic variables introducing a specific spacetime foliation. Restricting to a homogeneous and isotropic background, we perform a Legendre transformation in a subset of the perturbation variables reducing systematically the constraints.

The paper is organized as follows: in the next section we define some relevant geometrical objects and settle the notation and conventions. In Section~\ref{sec:GA} we obtain the second order gravitational Lagrangian for geometrical perturbations around a homogeneous and isotropic geometry with arbitrary spacelike hyper-surfaces, while in Section~\ref{sec:m:LA} we obtain the second order matter Lagrangian for arbitrary fluids. In Sections~\ref{sec:full:LA} and \ref{sec:full:HA} the full simplified second order action and Hamiltonian are obtained. We end up with the conclusions. In the Appendix~\ref{app:kpert} we obtain the relations between the tensor $\LL_{\mu\nu}{}^\beta$ and the perturbations on the kinetic variables and spatial curvature tensor. Then, using these relations, in the Appendix~\ref{sec:pert:L2} we obtain the second order Lagrangian in terms of these for a general background metric. Finally in the Appendix~\ref{sec:pert:Lm2} we express a general matter Lagrangian up to second order in the matter fields perturbations.

\section{Geometry and Spacetime Foliation}
\label{sec:three_one}

In this section we shall briefly define some relevant geometrical objects and fix our notation. The space-time Lorentzian metric $\g_{\mu\nu}$ has signature $(-1,1,1,1)$ and the covariant derivative compatible with this metric is represented by the symbol $\cd_\mu$, i.e., $\cd_\alpha \g_{\mu\nu} = 0$.

The Riemann tensor and its contractions are defined with the following convention
\begin{equation*}
2\cd_{[\mu}\cd_{\nu]}u_\alpha = \R_{\mu\nu\alpha}{}^\beta{}u_\beta,\quad  \R_{\mu\alpha} = \R_{\mu\nu\alpha}{}^\nu, \quad \R = \R_\mu{}^\mu,
\end{equation*}
where the symbols $[\;]$ and $(\;)$ represent the anti-symmetric and symmetric part of the tensor, i.e. $$ M_{[\mu\nu]} \equiv \frac{1}{2}\left(M_{\mu\nu} - M_{\nu\mu}\right), \quad M_{(\mu\nu)} \equiv \frac{1}{2}\left(M_{\mu\nu} + M_{\nu\mu}\right).$$

We define the foliation of the space-time through a normalized timelike vector field $\n^\mu$ normal to each spatial section ($\n^\mu\n_\mu = -1$). The foliation induces a metric in the hyper-surfaces as $\h_{\mu\nu} = \g_{\mu\nu} + \n_\mu\n_\nu$, which projects non-spatial vectors into the hyper-surfaces. For an arbitrary tensor $M_{\mu_1\dots\mu_m}^{\indspace\nu_1\dots\nu_k}$ the projector is defined as
$$\hp{M^{\indspace\nu_1\dots\nu_k}_{\mu_1\dots\mu_m}} \equiv \h_{\mu_1}^{\phantom a \alpha_1}\dots\h_{\mu_m}^{\phantom a\alpha_m}\h_{\beta_1}^{\phantom a \nu_1}\dots\h_{\beta_k}^{\phantom a\nu_k}M^{\indspace\beta_1\dots\beta_k}_{\alpha_1\dots\alpha_m}.$$
We shall call a spatial object, any tensor that is invariant under the projection, i.e. $\hp{M_{\mu_1\dots\mu_m}^{\indspace\nu_1\dots\nu_k}} = M_{\mu_1\dots\mu_m}^{\indspace\nu_1\dots\nu_k}$. The covariant derivative compatible with the spatial metric $\h_{\mu\nu}$ is
\begin{equation}
\scd_\alpha M_{\mu_1\dots\mu_m}^{\indspace\nu_1\dots\nu_k} = \hp{\cd_\alpha M_{\mu_1\dots\mu_m}^{\indspace\nu_1\dots\nu_k}}.
\end{equation}
The spatial covariant derivative compatible with $\h_{\mu\nu}$ defines the spatial Riemann curvature tensor
\begin{equation}\label{eq:def:SR}
[\scd_\mu\scd_\nu - \scd_\nu\scd_\mu]A_\alpha \equiv \SR_{\mu\nu\alpha}{}^\beta A_\beta,
\end{equation}
where $A_\beta$ is an arbitrary spatial field. The spatial Laplace operator is represented by the symbol $\scd^2$, i.e., $\scd^2 \equiv \scd_\mu\scd^\mu$, and we denote the contraction with the normal vector field $\n^\mu$ with an index $\n$, e.g., $M_{\alpha \n} \equiv M_{\alpha \beta}\n^\beta$.

The derivative of the velocity field defining the foliation can be decomposed as
\begin{equation}\label{eq:def:EC:2}
\cd_\mu\n_\nu = \EC_{\mu\nu} - \n_\mu\ac_\nu,
\end{equation}
with the acceleration and the extrinsic curvature defined, respectively, as $\ac_\mu \equiv \n^\gamma\cd_\gamma\n_\mu$ and $\EC_{\mu\nu} \equiv \hp{\cd_\mu\n_\nu}$. In addition, from the extrinsic curvature we define the expansion factor and the shear,\footnote{The Frobenius theorem guaranties that for a global spatial sectioning the normal field satisfy $\n_{[\alpha}\cd_{\mu}\n_{\nu]} = 0$, which can be expressed as $\cd_{[\mu}\n_{\nu]} = \ac_{[\mu}\n_{\nu]}$. Therefore, for a global spatial sectioning the vorticity is null, i.e., $\EC_{[\mu\nu]} = 0$.} i.e.,
\begin{equation}
\EX \equiv \EC_\mu{}^\mu, \qquad \SH_{\mu\nu} \equiv \EC_{\mu\nu} - \frac{\EX}{3}\h_{\mu\nu}.
\end{equation}

For a geodesic foliation, the Lie derivative of the projector is null ($\lie_\n\h_\mu{}^\nu = 0$). Therefore, in this case, the Lie derivative commutes with the projector, i.e.,
\begin{equation*}
\lie_\n\hp{M_{\mu_1\dots\mu_m}^{\indspace\nu_1\dots\nu_k}} = \hp{\lie_\n{M}_{\mu_1\dots\mu_m}^{\indspace\nu_1\dots\nu_k}}.
\end{equation*}

In what follows, we shall be interested in analyzing linear cosmological perturbations. Thus, it is imperative to properly define what it is meant by a metric perturbation. The space-time is defined by the physical metric $\g_{\mu\nu}$, which describes the actual physical dynamics. In addition, we shall suppose that exist a background metric $\bg_{\mu\nu}$ such that the physical metric $\g_{\mu\nu}$ can be seen as a ``small perturbation''. In other words, we suppose that the difference $\dg_{\mu\nu} = \g_{\mu\nu} - \bg_{\mu\nu}$ can be treated perturbatively in the sense discussed in~\cite{Vitenti2012}. Note, however, that one usually defines $\dg^{\mu\nu}$ as the difference between the inverse metric and its background value, i.e. $\dg^{\mu\nu} \equiv \g^{\mu\nu} - \bg^{\mu\nu}$, which, in general, is different from $\dg_{\alpha\beta}\bg^{\alpha\mu}\bg^{\beta\nu}$. Therefore, it is convenient to define the tensor $\ddg_{\mu\nu}$ and its covariant form as
\begin{align}\label{eq:def:ddg}
\ddg_{\mu\nu} \equiv \g_{\mu\nu} - \bg_{\mu\nu},\qquad \ddg^{\mu\nu}\equiv\bg^{\mu \alpha}\bg^{\nu\beta}\ddg_{\alpha \beta}.
\end{align}

The covariant derivative compatible with the background metric is represented by the symbol $\bcd$ or by a semicolon ``$\, ;\, $'', i.e. $\bg_{\mu\nu;\gamma} \equiv \bcd_\gamma\bg_{\mu\nu} = 0$. Using a background foliation described by the normal vector field $\bn^\mu$, we define the projector $\bh_{\mu\nu}$, spatial derivative $\bscd_\mu$ and spatial Riemann tensor $\bSR_{\mu\nu\alpha}{}^\beta$, as we have done for the objects derived from $\g_{\mu\nu}$. We use the symbol ``$ {}_{\scp} $'' to represent the background spatial derivative, i.e., $T_{\scp\mu} \equiv \bscd_\mu T$ for any tensor $T$.

One should keep in mind that the background tensors and the perturbation tensors have their indices lowered and raised always by the background metric. Finally, we define the dot operator of a arbitrary tensor as
\begin{equation}
\dot{M}_{\mu_1\dots\mu_m}^{\indspace\nu_1\dots\nu_k} \equiv \bhp{\lie_{\bn}{}M_{\mu_1\dots\mu_m}^{\indspace\nu_1\dots\nu_k}}.
\end{equation}

\section{Gravitational Action}
\label{sec:GA}

In what follows, we shall be concerned with the dynamics of linear cosmological perturbations. Accordingly, to obtain a system of first order dynamical equations, we must expand the Lagrangian up to second order. As it is well known, the gravitational part of the action is
\begin{equation}\label{eq:def:ACg}
\AC_\g = \int\dd^4x\LA_\g, \quad \LA_\g \equiv \frac{\sqrt{-\g}\R}{2\kp},
\end{equation}
with $\kp = 8\pi{}G/c^4$, $G$ the gravitation constant, $c$ the speed of light and $\R$ the curvature scalar. The expansion in the metric tensor Eq.~\eqref{eq:def:ddg} induces an expansion in the curvature tensor. A simple way to describe this expansion is through the difference between the covariant derivative of the perturbed and of the background metric. Accordingly, we define the true tensor $\LL_{\mu\nu}{}^\alpha$ by the equation
\begin{align}
&(\cd_\mu-\bcd_\mu)A_\nu = \LL_{\mu\nu}{}^\beta A_\beta,\nonumber\\
&\LL_{\alpha\beta}{}^\gamma = -\frac{\g^{\gamma\sigma}}{2}\left(\g_{\sigma\beta;\alpha} + \g_{\sigma\alpha;\beta} - \g_{\alpha\beta;\sigma}\right), \label{eq:LL}
\end{align}
and we remark that the covariant derivative ``$\, ;\, $'' is with respect to the background metric. Hence, in first order we have
\begin{equation}\label{eq:LL:1}
\LL_{\alpha\beta\gamma} = -\frac{1}{2}\left(\ddg_{\gamma\beta;\alpha} + \ddg_{\gamma\alpha;\beta} - \ddg_{\alpha\beta;\gamma}\right).
\end{equation}
We can also define two covectors as
\begin{align}
\LLa_\alpha &\equiv \LL_{\nu\alpha}{}^\nu = - \frac{\ddg_{;\alpha}}{2}, \\
\LLb_\mu &\equiv \bg^{\alpha\beta}\LL_{\alpha\beta\mu} = -\ddg_\mu{}^{\sigma}{}_{;\sigma} + \frac{\ddg_{;\mu}}{2},
\end{align}
and it is worth mentioning that the covectors above and the tensor $\LL_{\alpha\beta\gamma}$ are at least of first order in $\ddg_{\mu\nu}$. The perturbed Riemann tensor is related to the background Riemann tensor by
\begin{equation}\label{eq:pR:R}
\R_{\mu\nu\alpha}{}^\beta = \bR_{\mu\nu\alpha}{}^\beta + 2\LL_{\alpha[\nu}{}^\beta{}_{;\mu]} + 2\LL_{\alpha[\mu}{}^\gamma\LL_{\nu]\gamma}{}^\beta,
\end{equation}
where $\bR_{\mu\nu\alpha}{}^\beta$ is the Riemann tensor constructed with the background metric $\bg_{\mu\nu}$. Thus, the expansion of the curvature scalar up to second order is
\begin{equation}\label{eq:pSR:SR}
\begin{split}
\R &\approx \bR + \bR_{\mu\alpha}\dg^{\mu\alpha} + (\LLa^\mu - \LLb^\mu)_{;\mu} + \LLb^\mu\LLa_\mu \\
&- \LL_{\mu\nu\alpha}\LL^{\mu\alpha\nu} + (\LLa_{\nu;\mu}-\LL_{\mu\nu}{}^\gamma{}_{;\gamma})\dg^{\mu\nu},
\end{split}
\end{equation}
where again $\bR$ and $\bR_{\mu\alpha}$ are, respectively, the scalar curvature and the Ricci tensor of the background. To complete the expansion of the Lagrangian we also need the metric determinant up to second order, namely
\begin{equation}\label{eq:exp:sqrtmg}
\sqrt{-\g} \approx \sqrt{-\bg}\left[1 + \frac{\ddg}{2} - \frac{\ddg_{\mu\nu}\ddg^{\mu\nu}}{4} + \frac{\ddg^2}{8}\right].
\end{equation}

\subsection{Second Order Gravitational Lagrangian}

Using the above expansions [Eqs.~\eqref{eq:pSR:SR} and \eqref{eq:exp:sqrtmg}], we can decompose the gravitational Lagrangian given in Eq.~\eqref{eq:def:ACg} in powers of the perturbations $\ddg_{\mu\nu}$. Besides grouping the pure background part and the first order terms, we can distinguish in the second order Lagrangian a kinetic and a potential term. Thence, we decompose the second order expansion of the gravitational Lagrangian as
\begin{equation}
\LA_\g = \bLA_\g + \dLA_\g^{(1)} + \dLA_{\g\kin}^{(2)} + \dLA_{\g\pot}^{(2)} + \dLA_{\g\sur,1},
\end{equation}
where the background and the first order terms are, respectively,
\begin{align}
\bLA_\g = \sqrt{-\bg}\frac{\bR}{2\kp}, \quad \dLA_\g^{(1)}=-\frac{\sqrt{-\bg}}{2\kp}\bG_{\mu\nu}\ddg^{\mu\nu},
\end{align}
with $\bG_{\mu\nu}$ being the Einstein tensor evaluated with the background metric $\bg_{\mu\nu}$. These two terms are self-evident in view of the variational principle and need no further analysis. The last term $\dLA_{\g\sur,1}$ includes all surface terms
\begin{align}
\dLA_{\g\sur,1} &= \frac{\sqrt{-\bg}}{2\kp}\bigg(\LLa^\mu - \LLb^\mu+\LL_{\alpha\beta}{}^\mu\ddg^{\alpha\beta}\bigg)_{;\mu}\nonumber\\
&+ \frac{\sqrt{-\bg}}{2\kp}\bigg[(\LLa^\mu - \LLb^\mu)\frac{\ddg}{2} - \LLa_\nu \ddg^{\mu\nu}\bigg]_{;\mu},
\end{align}
and hence is irrelevant for dynamics. Finally, as we mentioned above, the second order Lagrangian can be split in a kinetic term $\dLA_{\g\kin}^{(2)}$ that includes derivatives of the perturbation and a potential term $\dLA_{\g\pot}^{(2)}$ without derivatives of the perturbation. These terms read
\begin{align}
\dLA_{\g\kin}^{(2)} &= \frac{\sqrt{-\bg}}{2\kp}\left[\LL^{\mu\nu\gamma}\LL_{\gamma(\mu\nu)} - \LLa_\mu\LLb^\mu\right],\label{eq:def:dLAk2}\\
\dLA_{\g\pot}^{(2)} &= \frac{\sqrt{-\bg}}{2\kp}\left(\bG_{\mu\nu}+\frac{\bg_{\mu\nu}}{4}\bR\right)\ddg^{\mu}{}_\alpha\left(\ddg^{\alpha\nu}-\frac{\bg^{\alpha\nu}\ddg}{2} \right).\label{eq:def:dLAp2}
\end{align}

Note that the kinetic term $\dLA_{\g\kin}^{(2)}$ is quadratic in $\LL_{\mu\nu}{}^\gamma$. Therefore, to express the Lagrangian in terms of the metric perturbation, we must first relate $\LL_{\mu\nu}{}^\gamma$ with the perturbed kinematic parameters associated with the spatial slicing. We should stress that this is a key step in our procedure. The conciseness of describing the metric perturbation in terms of the kinematic parameters is crucial to perform the very involved expansion of the gravitational Lagrangian without ever using the background equations. This calculation can be found in detail in Appendix~\ref{app:kpert} and~\ref{sec:pert:L2}. In Appendix~\ref{app:kpert}, we obtain the perturbations of the kinematic parameters for an arbitrary spatial slicing defined by the normal vector field $\n^\mu$, and in Appendix~\ref{sec:pert:L2} we relate them with the terms appearing in $\dLA_{\g\kin}^{(2)}$ for a generic metric.

In the case of a Friedmann background, the second order Lagrangian given in Eq.~\eqref{eq:final:dLA2} can be further simplified. For these metrics, there is a preferred slicing with normal vector $\bn^\mu$ in which the spatial sections are homogeneous and isotropic. Straightforward calculation shows that, for this foliation, the extrinsic curvature and the spatial Ricci tensor are diagonal $$\bEC_{\mu\nu} = \frac{\bEX}{3}\bh_{\mu\nu}, \qquad\bSR_{\mu\nu} = 2\bK\bh_{\mu\nu},$$ with the expansion factor and the function $\bK$ being homogeneous, i.e. $\bEX_{\scp\mu} = 0 = \bK_{\scp\mu}$. Thus, the Einstein tensor is given by
\begin{align}\label{eq:FLRW:ee}
\bG_{\mu\nu} &= \left(3\bK + \frac{\bEX^2}{3}\right)\bn_\mu\bn_\nu - \left(3\bK+2\dot{\bEX} + \bEX^2\right)\frac{\bh_{\mu\nu}}{3}.
\end{align}
The symmetries in the Friedmann metric simplify significantly the kinetic second order term in the gravitational Lagrangian. Given the background foliation $\bn^\mu$, the metric perturbation can be decomposed as
\begin{equation}
\ddg_{\mu\nu} = 2\A\bn_\mu\bn_\nu+2\B_{(\mu}\bn_{\nu)}+2\C_{\mu\nu},
\end{equation}
where (see Appendix \eqref{app:kpert} for details)
\begin{align*}
\A \equiv \frac12 \ddg_{\bn\bn}, \quad \B^\mu \equiv - \bhp{\ddg_{\bn}{}^{\mu}},\quad \C_{\mu\nu} \equiv \frac{\bhp{\ddg_{\mu\nu}}}{2}.
\end{align*}

Using the scalar, vector and tensor decomposition (see~\cite{Stewart1990}) we rewrite the metric perturbations as
\begin{align*}
\B_\mu &= \BS_{\scp\mu} + \BV_\mu, \\
\C_{\mu\nu} &= \CS\h_{\mu\nu} - \CSD_{\scp\mu\nu} + \CV_{(\nu\scp\mu)} + \CTD_{\mu\nu},
\end{align*}
where $\BV^\mu{}_{\scp\mu} = \CV^\mu{}_{\scp\mu} = \CTD_\mu{}^\nu{}_{\scp\nu} = \CTD_\mu{}^\mu = 0$. It is straightforward to show, using the results of Appendix~\ref{app:kpert}, that in terms of this decomposition the shear perturbation reads
\begin{equation}\label{eq:def:FLRW:sCI:decomp}
%\begin{split}
\dSH_{\mu\nu} = \left[\bscd_{(\mu}\bscd_{\nu)}-\frac{\bh_{\mu\nu}\blap}{3}\right]\dSHs + \dSHv_{(\nu\scp\mu)} + \dot{\CTD}_\mu{}^\alpha\bh_{\alpha\nu},
%\end{split}
\end{equation}
where we have defined
\begin{equation}\label{eq:FLRW:dSHs}
\dSHs \equiv \left(\BS-\dot{\CSD} + \frac{2}{3}\bEX\CSD\right), \quad \dSHv^{\alpha} \equiv \BV^\alpha + \dot{\CV}^\alpha.
\end{equation}
The perturbation on the expansion factor gives
\begin{equation}\label{eq:FLRW:dEX}
\dEX = \blap\dSHs  + \bEX\A + 3\dot{\CS}.
\end{equation}
Finally, the perturbations on the spatial Ricci tensor and curvature scalar are
\begin{align*}
\bhp{\dSR_\mu{}^\bn} &= 0, \\
\bhp{\dSR_\bn{}^\nu} &= -2\bK[\BS^{\scp\nu} + \BV^\nu], \\
\bhp{\dSR_\mu{}^\nu} &= -\CS_{\scp\mu}{}^{\scp\nu} - \bh_\mu{}^\nu[\blap + 4\bK]\CS - [\blap - 2\bK]\CTD_\mu{}^\nu, \\
\dSR &= -4\blapK\CS,
\end{align*}
where we have defined the operator $\blapK \equiv \blap + 3\bK$.

Using this decomposition, the second order gravitational Lagrangian is
\begin{equation}\label{eq:FLRW:dLA2:g}
\begin{split}
\frac{2\kp}{\sqrt{-\bg}}\dLA_{\g\kin}^{(2)} &= \dEC_\mu{}^\nu\dEC_\nu{}^\mu - \dEX^2 -\C_\mu{}^\nu\dSR_\nu{}^\mu \\
&+\left(\frac{\C}{2}- \A\right)\dSR + \bG_{\bn\bn}\left(\B_\mu\B^\mu-\A^2-2\A\C\right)\\
&+ \bG_{\mu\nu}\bh^{\mu\nu}\left(2\C^{\mu\alpha}\C_{\alpha\mu} - \C^2\right),
\end{split}
\end{equation}
with $\C \equiv \ddg_{\mu\nu}\bh^{\mu\nu} / 2$. In the above expression, the last two terms are proportional to the background Einstein tensor. Combining them with
the part coming from the matter Lagrangian, these terms will be proportional to the background Einstein equations. Thus, if the background equations are valid they vanish. However, we shall show that they can be eliminated by a simple redefinition of the perturbed fields, which is valid independently of the background equations. The main advantage is that the resulting Hamiltonian obtained without using the background equations can be used to not only quantize the perturbations, as it is commonly done in the literature, but also the background degrees of freedom.

\section{Perfect Fluids Action}
\label{sec:m:LA}

In reference~\cite{Schutz1970}, it has been shown that is possible to obtain the equations of motion for a perfect fluid from a variational principle. This formalism decomposes the four-velocity of the fluid $\ent_\mu$ in terms of potentials as
\begin{equation}\label{eq:def:ent:a}
\ent_\mu = \cd_\mu\fpa + \fpb\cd_\mu\fpc + \fpd\cd_\mu\en,
\end{equation}
where $\fp_{a}$ for $a=1,2,3,4$ are arbitrary scalar fields and $\en \equiv \fp_5$ represents the specific entropy of the fluid. In addition, the specific enthalpy is simply given by the module of the four-vector, i.e., $ \ent \equiv \sqrt{-\ent_\mu\ent_\nu \g^{\mu\nu}}$. For future use, it is convenient to define a normalized vector field
\begin{equation}\label{eq:def:nm}
\nm_\mu \equiv \frac{\ent_\mu}{\ent}.
\end{equation}

In principle, this vector-field can have non-zero vorticity, hence, it does not define a global foliation. Notwithstanding, this does not pose any problem since our procedure to study second order perturbations is still valid regardless of the existence of a global foliation. If eventually desired, one can always impose conditions on the scalar fields $\fp_{a}$'s to annihilate the vorticity and construct a cosmological model. Thus, we shall carry on the simplification scheme for the general case and only afterwards, when needed, we shall impose the conditions for a homogeneous and isotropic model.

The action for the fluid is
\begin{equation}\label{eq:def:ACm}
\AC_m = \int\dd^4x\LA_m,\quad \text{with} \qquad \LA_m = \sqrt{-\g}p(\ent,\en)
\end{equation}
where the pressure $p$ is a function of the enthalpy and of the entropy. In terms of the enthalpy, the first law of thermodynamics can be cast as
$$\dd\ent = \temp\dd\en + \frac{1}{\den}\dd\p, \quad\Rightarrow\quad \tdiff{\p}{\en}{\ent} = -\temp\den, \quad \tdiff{\p}{\ent}{\en} = \den.$$
Thus, first order variation of the Lagrangian with respect only to the fluid's degrees of freedom is
\begin{equation}
\delta \LA_m = -\sqrt{-\g}\left[\den\nm^\mu\delta\ent_\mu + \temp\den\delta\en\right].
\end{equation}
The perturbation in the vector field $\ent_\mu$ should be expressed in terms of the potentials $\fp_{a}$'s. Hence, by varying the action with respect to the $\fp_a$'s and $\en$, respectively, one has
\begin{equation}
\begin{split}\label{eq:bkg}
&\cd_\mu\left[\den\nm^\mu\right] = 0, \quad \cd_\nm{\fpc} = 0, \quad \cd_\nm{\fpb} = 0, \\
&\cd_\nm{\en} = 0, \quad \cd_\nm{\fpd} = \temp\den.
\end{split}
\end{equation}
Combining the above equations we obtain
\begin{equation}
\ent = \sqrt{-\ent^\mu\ent_\mu} = -\cd_\nm{\fpa}.
\end{equation}

Finally, the energy momentum tensor for the fluid is simply
\begin{equation}
\begin{split}
T_{\mu\nu} &= \den\ent\nm_\mu\nm_\nu + p \g_{\mu\nu} = \ed\nm_\mu\nm_\nu + p\h_{\mu\nu},
\end{split}
\end{equation}
where the energy density $\ed$ is defined by $\ed \equiv \den\ent - p$.

\subsection{Second Order Matter Lagrangian}

In this sub-section we shall perform the same expansion up to second order in the fluid Lagrangian as we have done for the gravitational part of the action. In the process of expanding the Lagrangian we shall need the following relations
$$\tdiff{\den}{\en}{\ent},\quad\tdiff{\den}{\ent}{\en},\quad\tdiff{\temp}{\ent}{\en}.$$
Recalling that $\ent = (\ed+\p)/\den$, we obtain
\begin{equation}\label{eq:def:ther12}
\begin{split}
&\tdiff{\den}{\ent}{\en} = \frac{\den}{\cs^2\ent}, \qquad \tdiff{\temp}{\ent}{\en} = \frac{\temp}{\ent} + \frac{\tal}{\den\ent\cs^2},\\
&\tdiff{\den}{\en}{\ent} = -\frac{\temp\den(1+\cs^2) + \tal}{\cs^2\ent}.
\end{split}
\end{equation}
with the speed of sound $\cs^2$ and $\tal$ defined as
$$\cs^2 \equiv \tdiff{\p}{\ed}{\en}, \quad \tal \equiv \tdiff{\p}{\en}{\ed}.$$

The first order term in the expansion of Eq.~\eqref{eq:def:ACm} is
\begin{equation}
\dLA_\mat^{(1)} = \wbackdec{\frac{\delta\AC_\mat}{\dmf_a}}\dmf_a + \frac{\sqrt{-\bg}}{2}\bT^ {\mu\nu}\ddg_{\mu\nu},
\end{equation}
where the overline means that the quantity should be evaluated using the background fields. To simplify the expressions, we define the notation representing, respectively, the perturbation of $F$ by varying only the matter fields or the metric as
\begin{equation}
\delta^{\mf}F \equiv \int\dd^4x\wbackdec{\frac{\delta{}F}{\dmf_a}}\dmf_a, \quad \delta^{\g}F \equiv \int\dd^4x\wbackdec{\frac{\delta{}F}{\delta\g_{\mu\nu}}}\dg_{\mu\nu},
\end{equation}
noting that one should sum over all scalar fields $\mf_a$'s. Thus, the second order part is
\begin{equation}\label{eq:def:ACm2}
\begin{split}
\frac{\dLA_\mat^{(2)}}{\sqrt{-\bg}} &= \left[\frac{\delta^{\mf}\T_{\mu\nu}}{2} + \frac{\delta^\g\T_{\mu\nu}}{4}\right]\ddg^{\mu\nu}-\frac{\bT_{\mu\nu}}{2}\left[\ddg^{\mu\alpha}\ddg_\alpha{}^\nu-\frac{\ddg\ddg^{\mu\nu}}{4}\right]\\
&+ \int\dd^4{}y\wbackdec{\frac{\delta^2\AC_\mat}{\dmf_a\dmf_b(y)}}\frac{\dmf_a\dmf_b(y)}{2\sqrt{-\bg}},
\end{split}
\end{equation}
where we wrote explicitly only the space-time coordinates that differs from $x$. The first two terms above are, respectively,
\begin{align}
\delta^\mf\T^{\mu\nu} &= \left[\bden\left(2\bh^{\alpha(\mu}\bn^{\nu)}-\bh^{\mu\nu}\bn^\alpha - \frac{\bn^\mu\bn^\nu}{\bcs^2}\bn^\alpha\right)\right]\delta\ent_\alpha \nonumber\\
&- \left(\frac{\btemp\bden+\btal}{\bcs^2}\bn^\mu\bn^\nu + \btemp\bden\bh^{\mu\nu}\right)\delta\en,\\
\delta^\g\T_\mu{}^{\nu} &= \bden\bent\frac{\A}{\bcs^2}\bn_\mu\bn^\nu + \bden\bent\bn_{\mu}\B^{\nu} + \bden\bent\A\bh_\mu{}^\nu,
\end{align}
while the last term gives
\begin{equation}
\begin{split}
\frac{1}{\sqrt{-\bg}} \int&\dd^4{}y\wbackdec{\frac{\delta^2\AC_\mat}{\dmf_a\dmf_b(y)}}\frac{\dmf_a\dmf_b(y)}{2} = \\
&=\frac{\bden}{2\bent}\left(\frac{\bn^\gamma\bn^\sigma}{\bcs^2}-\bh^{\gamma\sigma}\right)\delta\ent_\gamma\delta\ent_\sigma-\frac{\bden}{2}\delta^2\ent_\bn \\
&+\frac{\bqual}{2}\delta\en^2\frac{\btemp\bden(1+\bcs^2)+\btal}{\bcs^2\bent}\delta\en\delta\ent_\bn,
\end{split}
\end{equation}
with the second derivative of the pressure defined as
\begin{equation}\label{eq:def:ther4}
\qual \equiv \tdiff{^2\p}{\en^2}{\ent} = -\tdiff{(\temp\den)}{\en}{\ent}.
\end{equation}

In the above variational method, we have considered the enthalpy $\ent$ and the potentials $\mf_a$'s as the fundamental dynamical variables. However, it is convenient to rewrite the perturbative expansion in terms of the usual fluid variables, namely, the energy density, the enthalpy and the four-velocity of the fluid. To relate these quantities, we note that the first order perturbation of the enthalpy gives
\begin{equation}
\delta\ent = -\frac{\delta{}(\ent_\mu\ent_\nu\g^{\mu\nu})}{2\bent} = -\delta\ent_\mu\bn^\mu + \bent\A = -\delta\ent_\bn + \bent\A.
\end{equation}
In addition, the normalized velocity field $\nm_\mu$ Eq.~\eqref{eq:def:nm} is related to the background velocity field $\bn_\mu$ through
\begin{align}\label{eq:rel:nm:bn}
\nm_\mu &= \frac{\ent_\mu}{\ent} = \bn_\mu + \frac{\delta\ent_\mu}{\bent} - \frac{\bn_\mu}{\bent}\delta\ent\\
&= \bn_\mu - \A\bn_\mu + \vm_\mu, \quad \text{with}\quad \vm_\mu \equiv \bhp{\frac{\delta\ent_\mu}{\bent}}.\label{eq:def:fp:vu}
\end{align}
These expressions, together with the thermodynamic relations, allow us to rewrite the first order pressure and energy perturbations as
\begin{align}
\delta \wp &= \bcs^2\dved + \btal\delta\en,\\
\dved &= \frac{\bden}{\bcs^2}\delta\ent -\frac{\btemp\bden + \btal}{\bcs^2}\delta\en, \\
&= -\frac{\bden\delta\ent_\bn + (\btemp\bden + \btal)\delta\en - \bden\bent\A}{\bcs^2}.\nonumber
\end{align}

Combining all the above results we can write the second order Lagrangian for a general perfect fluid as
\begin{equation}\label{eq:def:dLAm2}
\begin{split}
&\frac{\dLA_\mat^{(2)}}{\sqrt{-\bg}} = \frac{\bden(\delta\ent_\bn)^2}{2\bcs^2\bent} - \frac{\bden\bent\vm_\gamma\vm^\gamma}{2} - \frac{\bden\delta^2{}\ent_\bn}{2} + \frac{\bqual(\dent)^2}{2} \\
&+ \frac{[\btemp\bden(1+\bcs^2)+\btal]\delta\en\delta\ent_\bn}{\bcs^2\bent}-\bden\bent\B^\gamma\vm_\gamma+ \C\delta\wp + \A\dved \\
&-\bden\bent\C\A - \frac{\bden\bent}{2\bcs^2}\A^2 +\frac{\bed}{2}\left(-\B_\gamma\B^\gamma + \A^2 + 2\C\A\right) \\
&+ \frac{\bp}{2}\left(-2\C_\gamma{}^\nu\C_\nu{}^\gamma + \C^2\right),
\end{split}
\end{equation}
where $\delta^2{}\ent_\bn \equiv 2(\dfpb\dot{\dfpc} + \dfpd\dot{\delta\en})$.

%%%%%%%%%%%%%%%%%%%%%%%%%%%%%%%%%%%%%%%%%%%%%%%%%%%%%%%%%%%%%%%%
%%%%%%%%%%%%%%%%%%%%%%%%%%%%%%%%%%%%%%%%%%%%%%%%%%%%%%%%%%%%%%%%
%%%%%%%%%%%%%%%%%%%%%%%%%%%%%%%%%%%%%%%%%%%%%%%%%%%%%%%%%%%%%%%%
%%%%%%%%%%%%%%%%%%%%%%%%%%%%%%%%%%%%%%%%%%%%%%%%%%%%%%%%%%%%%%%%
%%%%%%%%%%%%%%%%%%%%%%%%%%%%%%%%%%%%%%%%%%%%%%%%%%%%%%%%%%%%%%%%
%%%%%%%%%%%%%%%%%%%%%%%%%%%%%%%%%%%%%%%%%%%%%%%%%%%%%%%%%%%%%%%%
%%%%%%%%%%%%%%%%%%%%%%%%%%%%%%%%%%%%%%%%%%%%%%%%%%%%%%%%%%%%%%%%
\subsection{Homogeneous and Isotropic perfect fluid}
\label{sec:fluidLA}

In the last section, we have performed the expansion of the matter Lagrangian up to second order considering a general perfect fluid. Nevertheless, considering cosmological models with a FLRW metric, Einstein's equations require that the background energy and pressure of the fluid shall be homogeneous and isotropic. Even though this condition does not impose that all the potentials $\mf_a$'s are homogeneous and isotropic, we shall do so to avoid any entanglement in the perturbations such as mixing the vector and scalar sectors (see Appendix \eqref{sec:pert:Lm2} for details). Assuming that all the potentials are indeed homogeneous and isotropic, equations \eqref{eq:def:ent:a} and \eqref{eq:rel:nm:bn} give us
\begin{equation}\label{eq:def:decvu}
\vm_\mu = \bhp{\frac{\delta\ent_\mu}{\bent}}= \vms_{\scp\mu}, \;\;\; \vms \equiv \frac{\dfpa + \bfpb\dfpc + \bfpd\dent}{\bent}.
\end{equation}

Additionally, one has that the perturbation of $\denta_\bn=\denta_\mu \bn^\mu$ is given by
\begin{equation}\label{eq:FLRW:dentan}
\denta_\bn = \bent(\lie_\bn-\bcs^2\bEX)\vms -\btemp\dent + Q,
\end{equation}
where the function $Q$ has been defined as
\begin{equation}\label{eq:FLRW:dentaQ}
\begin{split}
Q &\equiv - \dfpc \BGE_\bfpb - \dent \BGE_\bfpd + \dfpb \BGE_\bfpc \\
&+ \left(\dfpd+\vms\tdiff{\bent}{\ben}{\bden}\right) \BGE_{\ben} + \vms\frac{\bcs^2\bent}{\bden} \BGE_{\bden},
\end{split}
\end{equation}
with the $E$'s representing the background equations of motion, i.e.
%where the classical background motion equations are defined as
\begin{equation}\label{eq:def:fluid:motion}
\begin{split}
&\BGE_\bfpb = \dot{\bfpb}, \quad \BGE_\bfpc = \dot{\bfpc}, \quad \BGE_{\ben} = \dot{\ben} \\
&\BGE_\bfpd = \dot{\bfpb} - \btemp,\quad \BGE_{\bden} = \dot{\bden} + \bEX\bden.
\end{split}
\end{equation}
The function $Q$ groups all the terms that are formed by perturbation variables multiplied by background equations~\eqref{eq:bkg}. Note that the Lagrangian Eq.~\eqref{eq:def:dLAm2} depends on $\dfpa$ only through $\vms$, i.e. we can make the change of variable $\dfpa\rightarrow \vms$ by inverting Eq.~\eqref{eq:def:decvu}. With this new variable we can rewrite the perturbed energy density as
\begin{equation}\label{eq:def:dved}
\dved= \ded - \frac{\bden Q}{\bcs^2}, \quad \ded \equiv -\frac{\bden\bent(\lie_\bn-\bcs^2\bEX)\vms + \btal\dent - \bden\bent\A}{\bcs^2},
\end{equation}
where we have defined the perturbed energy density $\ded$ by excluding all the terms that depend on the background equations. In the same way, we have for the pressure and particles density
\begin{align}
\delta\wp = \dpp  - \bden Q &,& \delta \aleph = \dden - \frac{\bden Q}{\bcs^2\bent},
\end{align}
where $\dpp$ and $\dden$ do not depend on the background equations and are given by
\begin{align}
\dpp &\equiv \bcs^2\ded + \btal\dent = -\bden\bent(\lie_\bn-\bcs^2\bEX)\vms + \bden\bent\A, \label{eq:def:dpp}\\
\dden &\equiv \frac{\ded - \btemp\bden\dent}{\bent} \nonumber\\
&= -\frac{\bden\bent(\lie_\bn-\bcs^2\bEX)\vms + \btal\dent - \bden\bent\A}{\bcs^2\bent} - \frac{\btemp\bden\dent}{\bent}.\label{eq:def:dden}
\end{align}

It is convenient to make some simplifications. Note that we can invert Eq.~\eqref{eq:def:dden} so that we have $\bent(\lie_\bn-\bcs^2\bEX)\vms$ in terms of $\delta n$ and $\delta s$, which then can be substituted in the Lagrangian.  In addition, we shall rewrite the terms involving $\C$ and $\vms_{\scp\gamma}$, respectively, as
\begin{align*}
\C\dpp - \bden\bent\C\A &= -\frac{\lie_\bn(\sqrt{-\bg}\bden\C\bent\vms)}{\sqrt{-\bg}} +\bden\C\vms\tdiff{\bent}{\ben}{\bden}\dot{\ben} \\
&+ \C\bent\vms(1+\bcs^2)(\lie_\bn+\bEX)\bden + \bden\bent\vms\dot{\C},
\end{align*}
and
\begin{align*}
\bden\bent\B^\gamma\vms_{\scp\gamma} &= (\bden\bent\B^\gamma\vms)_{\scp\gamma} - \bden\bent\B^\gamma{}_{\scp\gamma}\vms, \\
 \frac{\bden\bent\vms_{\scp\gamma}\vms^{\scp\gamma}}{2} &=  \left(\frac{\bden\bent\vms\vms^{\scp\gamma}}{2}\right)_{\scp\gamma} - \frac{\bden\bent\vms\blap\vms}{2}.
\end{align*}
We can use the thermodynamic relations [Eqs.~\eqref{eq:def:ther12} and \eqref{eq:def:ther4}] to write
\begin{align}\label{eq:def:ther5}
\tdiff{\btemp}{\ben}{\bden} &= \frac{1}{\bden}\left\{\frac{[\btemp\bden(1+\bcs^2)+\btal]^2}{\bcs^2\bden\bent}-\bqual\right\}, \\
\tdiff{\btemp}{\bden}{\ben} &= \frac{\btemp\bden\bcs^2 + \btal}{\bden^2}.
\end{align}

As a result, the Lagrangian Eq.~\eqref{eq:def:dLAm2} for a homogeneous and isotropic perfect fluid is given by
\begin{equation}\label{eq:def:dLA2:m:den}
\begin{split}
&\frac{\dLA_\mat^{(2)}}{\sqrt{-\bg}} = \frac{\bcs^2\bent}{2\bden}\dden^2 - \tdiff{\btemp}{\ben}{\bden}\frac{\bden\dent^2}{2} + \frac{\bden\bent\vms\blap\vms}{2} + l^\tbg_1 \\
&- \bden\left(\dfpd\dot{\dent} + \dfpb\dot{\dfpc} - \bent\dEX\vms + \bent\bEX\A\vms - \A\btemp\dent\right)\\
&- \frac{\bed}{2}\left(\B_\gamma\B^\gamma - \A^2 - 2\C\A\right) - \frac{\bp}{2}\left(2\C_\gamma{}^\nu\C_\nu{}^\gamma - \C^2\right),
\end{split}
\end{equation}
where the term $l^\tbg_1$ contains the quantities proportional to the background equations of motion, i.e.
\begin{equation}
\begin{split}
l^\tbg_1 &\equiv -Q\left(\frac{\delta\aleph+\dden}{2} + \bden\C\right) \\
&+ \vms\C\left[\bent(1+\bcs^2)\BGE_{\bden} +\bden\tdiff{\bent}{\ben}{\bden}\BGE_{\ben}\right].
\end{split}
\end{equation}

The kinetic term in the Lagrangian above is quadratic in $\dden$, which is not gauge invariant. The advantage of using a gauge invariant kinetic term is that the associated momentum will also be automatically gauge invariant. Hence, we will now eliminate $\dden$ from the above expressions. Note that
\begin{align}
\pvms &\equiv -\sqrt{-\bg}(\ded - \bEX\bden\bent\vms), \label{pvms}\\
\pdfpc &\equiv - \sqrt{-\bg}\bden\dfpb, \\
\pdent &\equiv -\sqrt{-\bg}\bden(\dfpd+\btemp\vms),
\end{align}
are gauge invariant combinations of the fluid variables. Given the definitions above and Eqs.~\eqref{eq:def:decvu}--\eqref{eq:def:dden}, we have
\begin{equation*}
\begin{split}
\frac{\bcs^2\bent}{2\bden}\dden^2 &= \frac{\bcs^2}{2\bden\bent}\frac{\pvms^2}{\sqrt{-\bg}^2} + \frac{\bden\bent\bcs^2\bEX^2}{2}\vms^2 + \frac{\bcs^2\btemp^2\bden + 2\btal\btemp}{2\bent}\dent^2 \\
&+ \A\bden(\bent\bEX\vms - \btemp\dent) - \bden\left(\bent\bEX\vms - \btemp\dent\right)\dot{\vms} \\
&- (\bcs^2\bden\btemp+\btal)\bEX\vms\dent.
\end{split}
\end{equation*}
In order to simplify the expression above, we rewrite the following two terms involving $\vms^2$ as
\begin{equation*}
\begin{split}
&\frac{\bden\bent\bcs^2\bEX^2}{2}\vms^2 - \bden\bent\bEX\vms\dot{\vms} = -\frac{\lie_\bn}{\sqrt{-\bg}}\left(\frac{\sqrt{-\bg}\bden\bent\bEX\vms^2}{2}\right) \\
&+ \left[\bent(1+\bcs^2)\BGE_{\bden} + \bden\tdiff{\bent}{\ben}{\bden}\BGE_{\ben}\right]\frac{\bEX\vms^2}{2} + \frac{\bden\bent\dot{\bEX}}{2}\vms^2,
\end{split}
\end{equation*}
and the two terms involving $\vms\dent$ as
\begin{equation*}
\begin{split}
&\bden\btemp\dent\dot{\vms} - [\bcs^2\bden\btemp+\btal]\bEX\vms\dent = \frac{\lie_\bn}{\sqrt{-\bg}}\left[\sqrt{-\bg}\bden\btemp\dent\vms\right]  - \bden\btemp\dot{\dent}\vms \\
&-\left\{\left[\btemp(1+\bcs^2)+\frac{\btal}{\bden}\right]\BGE_{\bden} + \tdiff{[\bden\btemp]}{\ben}{\bden}\BGE_{\ben}\right\}\dent\vms.
\end{split}
\end{equation*}
Using the two simplifications above and discarding the surface terms we obtain
\begin{equation}\label{eq:def:dLA2:m}
\begin{split}
&\frac{\dLA_\mat^{(2)}}{\sqrt{-\bg}} = \frac{\bcs^2}{2\bden\bent}\frac{\pvms^2}{\sqrt{-\bg}^2} + \Pben\frac{\dent^2}{2}  + \frac{\pdent \dot{\dent}}{\sqrt{-\bg}}  + \frac{\pdfpc \dot{\dfpc}}{\sqrt{-\bg}} \\
& + \frac{\bden\bent}{2}\vms\blapK\vms - \frac{3\kp}{4}(\bden\bent\vms)^2 + \bden\bent\dEX\vms + l^\tbg_1 + l^\tbg_2 \\
&- \frac{\bed}{2}\left(\B_\gamma\B^\gamma - \A^2 - 2\C\A\right) - \frac{\bp}{2}\left(2\C_\gamma{}^\nu\C_\nu{}^\gamma - \C^2\right),
\end{split}
\end{equation}
where the term multiplying $\dent^2$ is defined by
\[
\Pben \equiv \bden\frac{\bcs^2\btemp^2\bden + 2\btal\btemp}{\bent} - \bden\tdiff{\btemp}{\ben}{\bden}.
\]
Hence, $\dden$ has been eliminated from the kinetic term of matter Lagrangian. With these modifications the extra term including quantities proportional to the background equations is
\begin{equation}
\begin{split}
&l^\tbg_2 = -\frac{3\bden\bent}{4}\vms^2(\BGE_\bn + \BGE_{\bh}) \\
&-\left\{\left[\btemp(1+\bcs^2)+\frac{\btal}{\bden}\right]\BGE_{\bden} + \tdiff{[\bden\btemp]}{\ben}{\bden}\BGE_{\ben}\right\}\dent\vms \\
&+\frac{\bEX\vms^2}{2}\left[\bent(1+\bcs^2)\BGE_{\bden} +\bden\tdiff{\bent}{\ben}{\bden}\BGE_{\ben}\right],
\end{split}
\end{equation}
where $\BGE_\bn$ and $\BGE_{\bh}$ are the background Einstein's equations [see Eq.~\eqref{eq:FLRW:ee}], i.e.,
\begin{equation}
\BGE_\bn \equiv \bG_{\bn\bn} - \kp\bed, \quad \BGE_{\bh} \equiv \frac{\bG_{\mu\nu}\bh^{\mu\nu}}{3} - \kp\bp,
\end{equation}
and $$\BGE_\bn + \BGE_{\bh} = -\frac{2}{3}\left(\dot{\bEX} + \frac{3\kp\bden\bent}{2}-3\bK\right).$$

Similarly to Eq.~\eqref{eq:FLRW:dLA2:g} that describes the second order gravitational Lagrangian, the fluid Lagrangian has several terms proportional to the background equations given by $l^\tbg_1 + l^\tbg_2$. In addition, the last two terms of Eq.~\eqref{eq:def:dLA2:m} multiplying the background energy and pressure are identical to the terms appearing in \eqref{eq:FLRW:dLA2:g}. As we shall see in the next section, when these terms are combined with the ones coming from the gravitational sector, it is possible to discard them without ever using the background equations of motion.

%%%%%%%%%%%%%%%%%%%%%%%%%%%%%%%%%%%%%%%%%%%%%%%%%%%%%%%%%%%%%%%%%
%%%%%%%%%%%%%%%%%%%%%%%%%%%%%%%%%%%%%%%%%%%%%%%%%%%%%%%%%%%%%%%%%
%%%%%%%%%%%%%%%%%%%%%%%%%%%%%%%%%%%%%%%%%%%%%%%%%%%%%%%%%%%%%%%%%
%%%%%%%%%%%%%%%%%%%%%%%%%%%%%%%%%%%%%%%%%%%%%%%%%%%%%%%%%%%%%%%%%
%%%%%%%%%%%%%%%%%%%%%%%%%%%%%%%%%%%%%%%%%%%%%%%%%%%%%%%%%%%%%%%%%
\section{Full Second Order Lagrangian}
\label{sec:full:LA}

In the last two sections, we have expanded the gravitational and the matter Lagrangians separately. We shall now show that combining these results we can considerably simplify the total second order Lagrangian. For simplicity, in this section we will neglect the spatial surface terms which appear during the simplification. The full second order Lagrangian $\dLA^{(2)}\equiv \dLA_\g^{(2)} + \dLA_\mat^{(2)}$ is $$\dLA^{(2)}=\dLA^{(2,s)}+\dLA^{(2,v)}+\dLA^{(2,t)} + \sqrt{-\bg}\left[l^\tbg_1 + l^\tbg_2 + l^\tbg_3\right],$$ with the new background equations included in $l^\tbg_3$,
\begin{equation}
\begin{split}
l^\tbg_3 &= \frac{\BGE_\bn[\B_\gamma\B^\gamma - \A^2 - 2\C\A] + \BGE_{\bh}[2\C_\gamma{}^\nu\C_\nu{}^\gamma-\C^2]}{2\kp}.
\end{split}
\end{equation}
The extra indices indicate each one of the independent sectors of the linear perturbations, namely, the scalar,
\begin{equation}\label{eq:FLRW:dLA2:s}
\begin{split}
&\frac{\dLA^{(2,s)}}{\sqrt{-\bg}} = \frac{\blap\dSHs\blapK\dSHs}{3\kp} + \frac{\bcs^2}{2\bden\bent}\frac{\pvms^2}{\sqrt{-\bg}^2}  + \Pben\frac{\dent^2}{2}\\
& + \left(\frac{\CS}{2}-\A\right)\frac{\dSR}{2\kp} - \frac{1}{3\kp}\left(\dEX-\frac{3\kp\bden\bent\vms}{2}\right)^2 \\
&+ \bden\bent\frac{\vms\blapK\vms}{2} +\frac{\pdent\dot{\dent}}{\sqrt{-\bg}} + \frac{\pdfpc \dot{\dfpc}}{\sqrt{-\bg}} ,
\end{split}
\end{equation}
and the vector and tensorial sectors
\begin{align}\label{eq:FLRW:dLA2:v}
\frac{\dLA^{(2,v)}}{\sqrt{-\bg}} &= \frac{\dSHv_{(\alpha\scp\nu)}\dSHv^{(\alpha\scp\nu)}}{2\kp}, \\ \label{eq:FLRW:dLA2:t}
\frac{\dLA^{(2,t)}}{\sqrt{-\bg}} &= \frac{\dot{\CTD}_\nu{}^\gamma\dot{\CTD}_\gamma{}^\nu  + \CTD_\mu{}^\nu(\blap - 2K)\CTD_\nu{}^\mu}{2\kp}.
\end{align}

The Lagrangian above is already in a simplified form. However, as discussed in Sec.~\ref{sec:fluidLA} it is more convenient to have a gauge invariant kinetic term. To accomplish this, we note that the term $$\dEXi \equiv \dEX-\frac{3\kp\bden\bent\vms}{2} + \frac{3\dSR}{4\bEX},$$ is gauge invariant, while $$\CSI \equiv \CS - \frac{\bEX}{3}\dSHs,$$ is the usual gauge invariant variable defined in the literature. Rewriting the scalar Lagrangian using both variables one obtains
\begin{equation}\label{eq:FLRW:dLA2:s:gi}
\begin{split}
&\frac{\dLA^{(2,s)}}{\sqrt{-\bg}} = \frac{3\blap\CSI\blapK\CSI}{\kp\bEX^2} + \frac{\bcs^2}{2\bden\bent}\frac{\pvms^2}{\sqrt{-\bg}^2}  + \Pben\frac{\dent^2}{2}  - \frac{\dEXi^2}{3\kp} \\
& + \frac{9\bden\bent}{2\bEX^2}\CSvms\blapK\CSvms + \frac{\pdent\dot{\dent}}{\sqrt{-\bg}} + \frac{\pdfpc \dot{\dfpc}}{\sqrt{-\bg}},
\end{split}
\end{equation}
where we have defined the gauge invariant variable $\CSvms \equiv \CS + \bEX\vms/3$. In this manner, it appears in the total second order Lagrangian an additional term $$l^\tbg_4 \equiv -\frac{9}{8\kp\bEX^2}\left(\BGE_\bn+\BGE_{\bh}\right)\CS\dSR,$$ which comes from the following substitution
\begin{equation*}
\frac{3\dot{\CS}\dSR}{2\kp\bEX} = -\frac{\lie_\bn}{\sqrt{-\bg}}\left(3\frac{\sqrt{-\bg}\CS\dSR}{4\kp\bEX}\right)-\frac{\CS\dSR}{4\kp} + \frac{\dot{\bEX}\CS\dSR}{4\bEX^2}.
\end{equation*}

The $l^\tbg_3+l^\tbg_4$ terms contain the time-time and the space-space Einstein's equations, while the other two terms $l^\tbg_1+l^\tbg_2$ represent a combination of the fluid's background dynamics and Einstein's equations. Instead of assuming that these terms cancel out due to the background equations, we will follow another route.

Note that since we are interested in linear dynamical equations, our perturbative expansion has been truncated in second order. Furthermore, the expansion has been done using first order variables and keeping only terms quadratic in them.

Nonetheless, it is completely legitimate to define perturbed variables that already contain second order terms. In other words, suppose that we have a given second order combination of the perturbations, $\delta_2 f$. We can define, for instance, a new variable $\A_\text{new}$ such that $\A_\text{new} = \A + \delta_ 2 f$. In doing so, we guarantee that both variables agree at first order and become different only at second order. Hence, this kind of modification leaves the first order variables intact, but can modify the Lagrangian in second order without, however, modifying the equations of motion up to first order because we can assume that the zeroth order equations of motion are valid after variation of the Lagrangian. Briefly, such change of variables when inserted in
the Lagrangian does not modify the equations of motion up to first order in perturbation theory.

Let us explore this change of variables to simplify the second order Lagrangian. The first order total Lagrangian can be written as
\begin{equation}\label{eq:dLA1}
\frac{2\kp\dLA_\g^{(1)}}{\sqrt{-\bg}} = -\left(\bG^{\mu\nu} - \kp \bT^{\mu\nu}\right)\ddg_{\mu\nu} = -2\A \BGE_\bn - 2\C \BGE_{\bh}.
\end{equation}
Thus, if we make the change of variable
\begin{align}\label{eq:Atrans}
\A &\rightarrow \A - \frac{\B_\gamma\B^\gamma - \A^2 - 2\C\A}{2} + \frac{9\CS\dSR}{8\bEX^2}, \\ \label{eq:Ctrans}
\C_\mu{}^\nu &\rightarrow \C_\mu{}^\nu - \frac{2\C_\mu{}^\gamma\C_\gamma{}^\nu-\C\C_\mu{}^\nu}{2} + \frac{3\CS\dSR}{8\bEX^2}\bh_\mu{}^\nu,
\end{align}
all the term $l^\tbg_3 + l^\tbg_4$ will be canceled out. Additionally, we remove the remaining terms proportional to the Einstein's equations present in $l^\tbg_2$  with a second transformation
\begin{equation}\label{eq:ACtrans}
\A \rightarrow \A + \frac{3\kp\bden\bent\vms^2}{4}, \qquad \C_\mu{}^\nu \rightarrow \C_\mu{}^\nu + \frac{\kp\bden\bent\vms^2}{4}\bh_\mu{}^\nu.
\end{equation}
It is worth noting that the above transformations do not change the first order Lagrangian as long as the perturbed fields are modified only in second order. We can also perform the same procedure to eliminate the background equations for the fluid. Once more, we consider the total first order Lagrangian,
\begin{equation}\label{eq:dLAm1}
\begin{split}
\frac{\dLA_\mat^{(1)}}{\sqrt{-\bg}} &= -\frac{\lie_\bn(\sqrt{-\bg}\bden\bent\vms)}{\sqrt{-\bg}} + \bent\vms\BGE_{\bden} \\
&+ \bden(\dfpc\BGE_{\bfpb} - \dfpb\BGE_{\bfpc} + \dent\BGE_{\bfpd} - \dfpd\BGE_{\ben}),
\end{split}
\end{equation}
where we have used the expression for $\denta_\bn$ coming from Eq.~\eqref{eq:FLRW:dentan}. Therefore, one can check that the transformations that eliminate the term $l^\tbg_1$ are
\begin{equation}\label{eq:def:trans1}
\begin{split}
\vms &\rightarrow \vms - \bcs^2\vms\left(\frac{\delta\aleph+\dden}{2\bden}\right) + \vms\C, \\
\dfp_2 &\rightarrow \dfp_2 + \dfp_2\left(\frac{\delta\aleph+\dden}{2\bden} + \C\right),\\
\dfp_3 &\rightarrow \dfp_3 + \dfp_3\left(\frac{\delta\aleph+\dden}{2\bden} + \C\right), \\
\dfp_4 &\rightarrow \dfp_4 + \left(\dfp_4 + \tdiff{\bent}{\ben}{\bden}\vms\right)\left(\frac{\delta\aleph+\dden}{2\bden}\right) +\dfpd\C, \\
\dent &\rightarrow \dent + \frac{\dent}{\bden}\left(\frac{\delta\aleph+\dden}{2} + \C\right).
\end{split}
\end{equation}
Finally, the remaining terms in $l^\tbg_2$ are canceled by the following transformation
\begin{equation}
\begin{split}
\vms &\rightarrow \vms + \frac{\bEX(1+\bcs^2)\vms^2}{2} - \left[\frac{\btemp(1+\bcs^2)}{\bent}+\frac{\btal}{\bden\bent}\right], \\
\dfp_4 &\rightarrow \dfp_4 + \tdiff{\btemp}{\ben}{\bden}\dent\vms - \frac{\bEX\vms^2}{2}\tdiff{\bent}{\ben}{\bden}.
\end{split}
\end{equation}

Therefore, we have proved that one does not need to use the background equations to arrive at the standard second order Lagrangian. Indeed, by a well defined change of variables, the terms proportional to the background equations are eliminated. From a mathematical point of view, our result is definitely robust, but it remains the question of the physical meaning of these changes of variables. This issue can be settle by analyzing the definition of our fundamental perturbed fields.

Our basic perturbed variable $\ddg_{\mu\nu}$ has been defined using the difference between the physical metric and the fiducial metric with its co-indices, i.e. $\ddg_{\mu\nu} \equiv \g_{\mu\nu} - \bg_{\mu\nu}$. However, there is no reason to not choose the covariant difference between these metrics $\dg^{\mu\nu} \equiv \g^{\mu\nu} - \bg^{\mu\nu}$. Assuming $\dg^{\mu\nu}$ as the fundamental perturbed variable, its time-time component is written in terms of $\ddg_{\mu\nu}$, up to second order terms, as
\[
\bn_\mu\bn_\nu\dg^{\mu\nu} \approx -2\A - 4\A^2 - \B_\gamma\B^\gamma,
\]
which resembles the transformation Eq.~\eqref{eq:Atrans}. Note that a simple change of basic first order perturbed variables induces a second order change in the original variables in the same way as displayed by our transformations. Indeed, half of the terms in the transformation Eq.~\eqref{eq:Atrans} can be perfectly interpreted is this very same way.

Therefore, our result that the second order Lagrangian can be simplified by the kind of change of variables described above is an indication that neither $\dg_{\mu\nu}$ nor $\dg^{\mu\nu}$ should be viewed as the most convenient variables for cosmological perturbations. Indeed, there must exist a certain combination of them that implements the above transformations which simplifies significantly its physical description. Consequently, this combination should be viewed as the most convenient one.

%%%%%%%%%%%%%%%%%%%%%%%%%%%%%%%%%%%%%%%%%%%%%%%%%%%%%%%%%%%
%%%%%%%%%%%%%%%%%%%%%%%%%%%%%%%%%%%%%%%%%%%%%%%%%%%%%%%%%%%
%%%%%%%%%%%%%%%%%%%%%%%%%%%%%%%%%%%%%%%%%%%%%%%%%%%%%%%%%%%
%%%%%%%%%%%%%%%%%%%%%%%%%%%%%%%%%%%%%%%%%%%%%%%%%%%%%%%%%%%
%%%%%%%%%%%%%%%%%%%%%%%%%%%%%%%%%%%%%%%%%%%%%%%%%%%%%%%%%%%
%%%%%%%%%%%%%%%%%%%%%%%%%%%%%%%%%%%%%%%%%%%%%%%%%%%%%%%%%%%
\section{Second Order Hamiltonian}
\label{sec:full:HA}

In the last section we have obtained the full second order Lagrangian for the perturbations. However, in order to quantize such system one needs to perform the Legendre transformations to go to the Hamiltonian, and consequently we need the Poisson algebra of the dynamical variables.

A close inspection of Eq.~\eqref{eq:FLRW:dLA2:s} shows that the scalar Lagrangian does not depend on any of the time derivatives of the variables $(\A,\; \BS,\; \dmf_2,\; \dmf_4)$. On the other hand, it does depends linearly on time derivatives of $(\dmf_3,\; \dent)$ and quadratically on the time derivatives of $(\CS,\; \CSD,\; \vms)$. Therefore, the Hessian matrix is singular and, as expected, we have a constrained Hamiltonian system.

Here we shall develop the Hamiltonian formalism only for the scalar sector of the simplified second order Lagrangian since it is the most involved one, and is more connected to cosmological observations. The appropriate generalizations to the vector and tensorial sectors are straightforward.

Instead of using Dirac's formalism~\cite{Dirac1931,Dirac1949}, we can deal with the constraints by using the procedure developed in Refs.~\cite{Faddeev1988,Jackiw1993}. These two formalisms are equivalent but the latter is operationally simpler than Dirac's, inasmuch some of the constrains are solved during the transitions to the Hamiltonian description. In this way, we keep only the necessary degrees of freedom.

The quadratic term in the Lagrangian depends on time derivative of $\CS$ and $\CSD$ only through $\dSHs$ and $\dEX$, Eqs.~\eqref{eq:FLRW:dSHs}-\eqref{eq:FLRW:dEX}. However, $\dEX$ depends on both $\dot{\CS}$ and $\dot{\CSD}$. Therefore, is simpler to use the variable $$\CST \equiv \frac{\C}{3} = \CS - \frac{\blap\CSD}{3},$$ instead of $\CS$, since by using this variable the perturbation on the expansion tensor is $$\dEX = 3\dot{\CST} + \blap\BS + \A\bEX.$$

The first step is to reduce Eq.~\eqref{eq:FLRW:dLA2:s:gi} to a Lagrangian linear in time derivatives by performing a Legendre transformation on the variables $(\CST,\; \blap\CSD,\; \vms)$. Thus, we shall perform the Legendre transformations only in those variables that we can invert their relations with the momenta. Accordingly, we define their momenta as
\begin{align}\label{eq:def:pCSD}
\pCSD &\equiv \bilapK\frac{\del\dLA^{(2,s)}}{\del{\lie_\bn[\blap\CSD]}} = \frac{2\sqrt{-\bg}}{\kp\bEX}\CSI, \\
\pCST &\equiv \frac{\del\dLA^{(2,s)}}{\del\dot{\CST}} = - 2\sqrt{-\bg}\frac{\dEXi}{\kp},
\end{align}
and $\pvms= \frac{\del\dLA^{(2,s)}}{\del\dot{\vms}}$ has already been defined in Eq.~\eqref{pvms}. The next step is to invert these relations to obtain the time derivatives in terms of the momenta,
\begin{align}
&\lie_\bn[\blap\CSD] = \blap\BS + \frac{3\kp}{2\sqrt{-\bg}}\blap\pCSD - \frac{3\blap\CS}{\bEX}, \\
&\dot{\CST} = \frac{\kp\bden\bent}{2}\vms-\frac{\bEX}{3}\A - \frac{\blap\BS}{3} + \frac{\blapK\CS}{\bEX}- \frac{\kp\pCST}{6\sqrt{-\bg}}, \\
&\dot{\vms} = \frac{\bcs^2\pvms}{\sqrt{-\bg}\bden\bent} + \A - \frac{\btal}{\bden\bent}\dent.
\end{align}
With these variables, the Lagrangian reads
\begin{equation}\label{eq:La:const}
\begin{split}
\dLA^{(2,s)} &= \blapK\pCSD\lie_\bn{\blap\CSD} + \pCST\dot{\CST} + \pvms\dot{\vms} + \pdent\dot{\dent} \\
& + \pdfpc \dot{\dfpc}  - \dHA_c^{(2,s)},
\end{split}
\end{equation}
where the constrained Hamiltonian $\dHA_c^{(2,s)}$ is
\begin{equation}\label{eq:Ha:const}
\begin{split}
&\dHA_c^{(2,s)} = \frac{\bcs^2\pvms^2}{2\sqrt{-\bg}\bden\bent} + \frac{3\kp\blap\pCSD\blapK\pCSD}{4\sqrt{-\bg}}  - \frac{\kp\pCST^2}{12\sqrt{-\bg}} \\
&+ \pCST\left(\frac{\kp\bden\bent\vms}{2}+\frac{\blapK\CS}{\bEX}\right) - \frac{\btal\pvms\dent}{\bden\bent} - \frac{3\blapK\pCSD\blap\CS}{\bEX} \\
&-\sqrt{-\bg}\left(\frac{9\bden\bent}{2\bEX^2}\CSvms\blapK\CSvms + \Pben\frac{\dent^2}{2}\right) \\
&+ \A\left(\pvms-\frac{\bEX\pCST}{3}\right) + \blap\BS\left(\blapK\pCSD -\frac{\pCST}{3} \right).
\end{split}
\end{equation}

In this intermediary stage we have a Lagrangian that is linear on any time derivative and is a function of seven variables $\left(\CST,\; \blap\CSD,\; \vms,\; \dent,\; \dmf_3,\; \A,\; \BS\right)$ but only of five momenta $\left(\pCST,\; \pCSD,\; \pvms,\; \pdent,\; \pdfpc \right)$. The extra two variables, namely $(\A,\;\BS)$ have no time derivative and the Lagrangian is only linear in these terms. Thus, they can be safely treated just as Lagrange multipliers. By varying the Lagrangian with respect to $(\A,\;\BS)$ one obtains
\begin{align}\label{eq:constraints}
\pvms - \frac{\bEX}{3}\pCST = 0 \quad \text{and} \quad \blapK\pCSD -\frac{\pCST}{3} = 0.
\end{align}

The above constraint equations show that from the three momenta $(\pCST,\; \pCSD,\; \pvms)$, there is actually only one linearly independent variable. Therefore, we can rewrite all of them in terms of just one. The choice is irrelevant, and leads to the same result. Nevertheless, there is a shorter route, which is to privilege $\pCSD$. Thence, let us write $\pCST$ and $\pvms$ in terms of $\pCSD$ as $$\pCST = 3\blapK\pCSD,\qquad \pvms = \bEX\blapK\pCSD.$$

We can now use this result to simplify the Lagrangian. In terms of only one momentum the Lagrangian reads
\begin{equation*}
\dLA^{(2,s)} = 3\blapK\pCSD\dot\CSvms + \pdent\dot{\dent} + \pdfpc \dot{\dfpc}  - \dHA^{(2,s)},
\end{equation*}
where we have dropped the subindex `$c$' in the Hamiltonian since we have solved all the constraints. A direct calculation shows that the unconstrained Hamiltonian is given by $\dHA^{(2,s)} =  \dHA_c^{(2,s)} + \dot{\bEX}\pvms\vms$, with all momenta written in terms of $\pCSD$. Hence,
\begin{equation*}
\begin{split}
&\dHA^{(2,s)} = \frac{\bcs^2\bEX^2\blap\pCSD\blapK\pCSD}{2\sqrt{-\bg}\bden\bent}- \frac{\btal\bEX\blapK\pCSD\dent}{\bden\bent} \\
&+\left(3\bK- \frac{3\kp\bden\bent}{2}+\bcs^2\bEX^2\right)\frac{3\bK\pCSD\blapK\pCSD}{2\sqrt{-\bg}\bden\bent} \\
&-\sqrt{-\bg}\left(\frac{9\bden\bent}{2\bEX^2}\MS\blapK\MS + \Pben\frac{\dent^2}{2}\right) - \frac{3}{2}\left(\BGE_\bn + \BGE_{\bh}\right)\vms\blapK\pCSD,
\end{split}
\end{equation*}
where we have defined a new variable
\begin{equation}\label{def:MS}
\MS \equiv \CSvms-\frac{\bK\bEX\pCSD}{\sqrt{-\bg}\bden\bent} = \CSvms - \frac{2\bK}{\kp\bden\bent}\CSI.
\end{equation}

The variable $\MS$ was introduced just to complete the square and hence to avoid any cross term between $\CSvms$ and $\pCSD$. However, the kinetic part of the Lagrangian is still written in terms of $\CSvms$ variable. By rewriting this term, we obtain
\begin{equation*}
\begin{split}
&3\blapK\pCSD\dot\CSvms = 3\blapK\pCSD\dot{\MS} + \lie_\bn{\left(\frac{3\bK\bEX\pCSD\blapK\pCSD}{2\sqrt{-\bg}\bden\bent}\right)} \\
&-\left(\frac{\BGE_{\bden}}{\bden}+\frac{\bcs^2\dot{\bden}}{\bden}+\tdiff{\bent}{\ben}{\bden}\dot{\ben}-
\frac{\dot{\bEX}}{\bEX}\right)\frac{3\bK\bEX\pCSD\blapK\pCSD}{2\sqrt{-\bg}\bden\bent}.
\end{split}
\end{equation*}
Therefore, discarding a surface term, the Hamiltonian in terms of $\MS$ simplifies to
\begin{equation}\label{def:dHA2s}
\begin{split}
\dHA^{(2,s)} &= \frac{\bcs^2\bEX^2\blap\pCSD\blapK\pCSD}{2\sqrt{-\bg}\bden\bent}- \frac{\btal\bEX\blapK\pCSD\dent}{\bden\bent} \\
&-\sqrt{-\bg}\left(\frac{9\bden\bent}{2\bEX^2}\MS\blapK\MS + \Pben\frac{\dent^2}{2}\right).
\end{split}
\end{equation}

As a result of this substitution, the Lagrangian gains an additional term $l^\tbg_5$ that has only terms proportional to the background equations, i.e.
\begin{equation*}
\begin{split}
l^\tbg_5 &= -\left[(1 + \bcs^2)\frac{\bEX\BGE_{\bden}}{\bden}+\tdiff{\bent}{\ben}{\bden}\bEX\BGE_{\ben}\right]\frac{3\bK\pCSD\blapK\pCSD}{2\sqrt{-\bg}\bden\bent} \\
&- \frac{3}{2}\left(\BGE_\bn + \BGE_{\bh}\right)\left(\frac{3\bK\pCSD\blapK\pCSD}{2\sqrt{-\bg}\bden\bent}-\vms\blapK\pCSD\right).
\end{split}
\end{equation*}
As the final step, we can perform another change of variables in order to cancel the term $l^\tbg_5$. One can check that the adequate transformation is
\begin{equation}
\begin{split}
\vms &\rightarrow \vms - \frac{3\bK(1+\bcs^2)\bEX\pCSD\blapK\pCSD}{2(\sqrt{-\bg}\bden\bent)^2}, \\
\dfp_4 &\rightarrow \dfp_4 + \tdiff{\bent}{\ben}{\bden}\frac{3\bK\bEX\pCSD\blapK\pCSD}{2(\sqrt{-\bg}\bden)^2\bent}, \\
\A &\rightarrow \A + \frac{3}{2}\left(\frac{3\kp\bK\pCSD\blapK\pCSD}{2\sqrt{-\bg}^2\bden\bent}-\vms\blapK\pCSD\right), \\
\C &\rightarrow \C + \frac{3}{2}\left(\frac{3\kp\bK\pCSD\blapK\pCSD}{2\sqrt{-\bg}^2\bden\bent}-\vms\blapK\pCSD\right).
\end{split}
\end{equation}

Therefore, we have that the second order Lagrangian to be used in the variational principle is
\begin{equation}\label{eq:dLA:s:uc}
\dLA^{(2,s)} = 3\blapK\pCSD\dot{\MS} + \pdent\dot{\dent} + \pdfpc \dot{\dfpc}  - \dHA^{(2,s)},
\end{equation}
with the second order Hamiltonian $\dHA^{(2,s)}$ given by Eq.\eqref{def:dHA2s}.

It is worth emphasizing the steps made up to this point. When the constraints [Eqs.~\eqref{eq:constraints}] are solved, the variable
\begin{equation}
\CSvms = \CS + \frac{\bEX\vms}{3},
\end{equation}
appears naturally as the combination of the perturbations involving a time derivative. However, in terms of this variable, the Hamiltonian contains cross terms of $\CSvms$ and its momentum $3\blapK\pCSD$. To avoid these terms, we defined a new variable $\MS$ in which the Hamiltonian is diagonal with respect to $\MS$ and its momentum:
\begin{equation}
\pMS \equiv 3\blapK\pCSD = \frac{6\sqrt{-\bg}}{\kp\bEX}\blapK\CSI.
\end{equation}

To express the results above in a more familiar manner, we first change the time derivative to a time defined by the congruence $l^\mu = N\n^\mu$. We denote the time derivatives with respect to $l^\mu$ as $T^\prime \equiv \lie_l T = N\dot{T}$, where the last equality is valid only when $T$ is a scalar. In terms of this time variable the Lagrangian reads
\begin{equation}\label{eq:LA:final}
\dLA^{(2,s)} = \pMS\MS^\prime + \pdent\dent^\prime + \pdfpc\dfpc^\prime  - \dHA^{(2,s)},
\end{equation}
where the Hamiltonian is given by
\begin{equation}\label{def:dHA2ss}
\begin{split}
\dHA^{(2,s)} &= N\Bigg[\frac{\kp\pMS\blap\bilapK\pMS}{4N\zz^2}- \frac{\kp\uu^2\pMS\dent}{2N\zz^2} \\
&-\left(\frac{N\bcs^2\zz^2}{\kp}\MS\blapK\MS + \sqrt{-\bg}\Pben\frac{\dent^2}{2}\right)\Bigg],
\end{split}
\end{equation}
and we defined the background quantities
\begin{align}
\zz &\equiv \frac{3}{\bcs\bEX}\sqrt{\frac{\sqrt{-\bg}\kp\bden\bent}{2N}} = \frac{N}{\bcs{}\gH}\sqrt{\frac{a^3\kp(\bed+\bp)}{2N}}, \\
\uu &\equiv \sqrt{\frac{3\sqrt{-\bg}\btal}{\bEX\bcs^2}} = \sqrt{\frac{Na^3\btal}{\gH\bcs^2}},
\end{align}
Note that $N\zz^2$ does not depend on $N$ and, therefore, Eq.~\eqref{def:dHA2ss} is linear on the lapse function. In the last equality above, we defined the Hubble function $H = \bEX / 3$ and the scale factor $a$, with $H = \dot{a}/a$. With these definitions, one can write $\sqrt{-\bg} = a^3$. For a general time variable, we express the Hubble function as $\gH \equiv a^\prime/a = NH$. It is worth noting that, for a conformal time ($N = a$), one obtains the same expression for $\zz$ as in the literature (see Eq.~(10.43b) in~\cite{Mukhanov1992}). From the Lagrangian in Eq.~\eqref{eq:LA:final}, we read the following Poisson structure:
\begin{equation}\label{eq:def:poisson:D}
\begin{split}
\Poisson{\MS(x)}{\pMS(y)} &= \dirac{3}{x-y}, \\
\Poisson{\dent(x)}{\pdent(y)} &= \dirac{3}{x-y}, \\
\Poisson{\dfpc(x)}{\pdfpc(y)} &= \dirac{3}{x-y},
\end{split}
\end{equation}
where $x$ and $y$ are coordinates at the same spatial hyper-surface and all other Poisson brackets are equal to zero. Thus, the time derivative of any perturbation field $A$ is given by $$A^\prime = \Poisson{A}{\int\dd^3x\dHA^{(2,s)}}.$$

We can also solve the momentum $\pMS$ as a function of $\MS^\prime$. Varying the Lagrangian with respect to $\pMS$ we obtain
\begin{equation}
\pMS = \blapK\bilap\left(\frac{2\zz^2}{\kp}\MS^\prime + \uu^2\dent\right).
\end{equation}
Substituting back in the Lagrangian of Eq.~\eqref{eq:LA:final}, one obtains
\begin{equation*}
\begin{split}
\dLA^{(2,s)} &= \frac{\ms^\prime\blapK\bilap\ms^\prime}{2} + \frac{\ms}{2}\blapK\left(\frac{\zz^{\prime\prime}}{\zz}\bilap+N^2\bcs^2\right)\ms\\
&+ \pdent\dent^\prime + \pdfpc\dfpc^\prime - \sqrt{\frac{\kp}{2}}\frac{\ms}{\zz}\blapK\bilap(\uu^2\dent)^\prime \\
& + \left(N\sqrt{-\bg}\Pben+\frac{\kp\uu^4}{2\zz^2}\right)\frac{\dent^2}{2},
\end{split}
\end{equation*}
where we have discarded the surface terms. In the Lagrangian above, we introduced the Mukhanov-Sasaki variable $\ms \equiv \MS\zz\sqrt{2/\kp},$ which coincide with the one defined in Eq.~(10.61) of Ref.~\cite{Mukhanov1992}. Note, however, that the Lagrangian stated in this reference [Eq.~(10.62)] is not valid for a non spatially flat background metric.

Finally, the total Hamiltonian up to second order is
\begin{equation}
\label{ham27}
\mathbf{H} = \mathbf{H}^{(0)} + \int\dd^3x\dHA^{(2,s)},
\end{equation}
where the zero order Hamiltonian reads
\begin{equation}
\label{ham270}
\mathbf{H}^{(0)} = N\left(\cV a^3\bed - \frac{3\cV a\cK}{\kp} - \frac{\kp\pa^2}{12\cV a}\right),
\end{equation}
$\dHA^{(2,s)}$ is given in Eq.~\eqref{def:dHA2ss}, and $\cV$ is the comoving volume and $\cK \equiv a^2\bK$ the comoving spatial curvature. In this construction, the energy density is $\bed \equiv \bed(\bden,\ben),$ where $\bden \equiv -\pbfpa/(\cV a^3)$ and the Poisson structure is given by the expressions
\begin{align*}
\Poisson{a}{\pa} &= 1,\quad \Poisson{\bfpa}{\pbfpa} = 1,\\
\Poisson{\bfpc}{\pbfpc} &= 1, \quad \Poisson{\ben}{\pben} = 1.
\end{align*}

Accordingly, we have explicitly obtained the second order Hamiltonian and written the complete Hamiltonian including its zero order part. However, in general, besides these terms, the full action still contains the first order Lagrangian, i.e. the $\dLA^{(1)}$. To deal with these linear terms, one can use different approaches. For example, in~\cite{Halliwell1985} the authors define a first order Hamiltonian and argue that in a perturbative regime they are not relevant to the quantization inasmuch as they can be modified by changing the ordering of the operatores and eventually only add a global phase to the quantum state. Alternatively, in a classical Friedmann model, by redefining the background variables one can always make each perturbed field to have zero spatial mean, which discard the first order Lagrangian all together (see for instance~\cite{Pinho2007}). 
%At present, we leave this issue aside to be carefully treated in a future work where we will discuss the actual simultaneous quantization of both background and perturbations.

Note that our calculations have been done as general as possible, namely, for an arbitrary spatial curvature $\K$ and for an arbitrary fluid, which can include entropy perturbations. Particularly important is the fact that we have nowhere used the background equations of motion. Thus, this Hamiltonian system for the second order fields can be simultaneously quantized with the background degrees of freedom.

%%%%%%%%%%%%%%%%%%%%%%%%%%%%%
%%%%%%%%%%%%%%%%%%%%%%%%%%%%%
%%%%%%%%%%%%%%%%%%%%%%%%%%%%%
%%%%%%%%%%%%%%%%%%%%%%%%%%%%%
\section{Comparison of the methods}

We have repeatedly stressed that our simplification procedure using field redefinitions eliminates terms proportional to the background equations, but it is important to remark that this is not equivalent to use these equations explicitly.

As a matter of fact, the first order Lagrangian is composed of terms that are always in the form of a zero order equation of motion multiplying a single perturbed field. This is essentially a consequence of the fact that the equations leave the action stationary.

As a result, any redefinition of the perturbed variables including second order terms will inevitably introduce terms proportional to the background field equations in the second order Lagrangian. These are exactly the extra terms that we have used to simplify the second order Lagrangian. Notwithstanding, one should note that we are only allowed to implement contact transformations, as you are working in the Lagrangian formalism, which limits the types of terms we can eliminate in the second order Lagrangian. Thus, it is not a priori evident that one can fully simplify the second order action through these transformations, but one has indeed to show case by case that all unwanted terms can indeed be discarded. As an example, we mention that any second order term involving time derivatives of the perturbations cannot be eliminated. Therefore, terms in the second order Lagrangian that are proportional, for instance, to $\BGE_{\bed}\vms \dot{\vms}$ (defined below) cannot be discarded through our transformations.

Another very important point is that the use of the background field equations produces an ambiguity in the form of the second order Lagrangian and, for this reason, an ambiguity in the perturbed momenta obtained from it. To be more specific, let us consider the momentum ${\pvms}$ that is defined in Eq.~\eqref{pvms} as  $\pvms = -\sqrt{-\bg}  (\ded - \bEX\bden\bent\vms)$. We can easily identify that the term multiplying $\vms$ is half of the background energy density conservation equation [see Eqs.~\eqref{eq:def:fluid:motion}], i.e. $\BGE_{\bed}=0$, where $$\BGE_{\bed} \equiv \bent\BGE_{\bden} + \btemp\bden\BGE_{\ben} = \dot{\bed}+\bEX\bden\bent = \dot{\bed} + \bEX(\bed+\bp).$$

Suppose now that during our procedure, instead of the above expression defined in Eq.~\eqref{pvms}, we had bumped into a similar term such as $\pvms = -\sqrt{-\bg}(\ded + \dot{\bed}\vms)$. At the classical level, or when quantizing only the perturbations, where the background equations are valid, these two expressions are equivalent. Nevertheless, this equivalence is not fulfilled at full quantum level, where quantum expectation values of physical quantities like $\ded$ may differ, as its relation to $\pvms$ is not the same. Therefore it seems imperative to discern between these two definitions of the ${\pvms}$ momentum.

To change the definitions of ${\pvms}$ as above one has to use the $\BGE_{\bed}$ term, as we can write ${\pvms}/ \sqrt{-\bg}=-(\ded + \dot{\bed}\, \vms) + \BGE_{\bed} \vms$. As long as the momentum $\pvms$ appears squared in the second order Lagrangian, the change of definitions of this momentum is equivalent to have two extra terms in the Lagrangian one being proportional to $\BGE_{\bed}\ded \, \vms\propto \BGE_{\bed}\vms\dot{\vms} $, which is exactly the type of term that our procedure cannot eliminate. Therefore, this ambiguity in the definition of $\pvms$ is not present in our procedure, contrary to the situation where one assumes the background field equations. As a matter of fact, following our procedure all the way through, gives us a unique and unambiguous definition of the momenta.

The discussion above revolves about the Lagrangian action principle, therefore, one could wonder whether there are transformations in the Hamiltonian formalism capable of dealing with the terms discussed above. It is easy to check that the type 2 canonical transformation applied to the constrained Hamiltonian in Eq.~\eqref{eq:Ha:const}, with the generator $$G_2(\vms,\pvms^\text{new}) = \sqrt{-\bg}\BGE_{\bed}\frac{\vms^2}{2} + \vms\pvms^\text{new},$$ transforms the momentum as $$\pvms \rightarrow \pvms^\text{new} = \pvms - \sqrt{-\bg}\BGE_{\bed} \vms = -\sqrt{-\bg}(\ded + \dot{\bed}\vms).$$ However, this transformation is time dependent and, for this reason, the Hamiltonian should transform as $$\dHA_c^{(2,s)} \rightarrow \dHA_c^{(2,s)} + \left(\bEX\sqrt{-\bg}\BGE_{\bed}+\sqrt{-\bg}\dot{\BGE}_{\bed}\right)\frac{\vms^2}{2},$$ introducing a time derivative of the equations of motion $\BGE_{\bed}$, which cannot be removed from the Hamiltonian using transformations involving the perturbative first order variables. This is similar to what happens in the Lagrangian formalism, using contact transformations: one could try to eliminate the term $\sqrt{-\bg}\BGE_{\bed}\vms\dot{\vms}$ rewriting it as $$\lie_{\bn}\left(\sqrt{-\bg}\BGE_{\bed}\frac{\vms^2}{2}\right) - \left(\bEX\sqrt{-\bg}\BGE_{\bed}+\sqrt{-\bg}\dot{\BGE}_{\bed}\right)\frac{\vms^2}{2},$$ and then remove the additional terms using a contact transformation. However, in this case, these terms are not proportional to the zeroth order equations of motion and, hence, they cannot be removed with the methods we have been using so far.

Summarizing, there are at least two strong reasons to implement our procedure instead of using the background equations of motion. The first reason stems from the fact that the change of variables technique implemented in this article may not yield, in general, the same second order Lagrangian. One has to show that all the terms to be simplified are indeed of the allowed form. Secondly, and most important, our procedure solves the ambiguity in the definition of the perturbed momenta if one assumes the validity of the background field equations. Hence, in order to obtain the correct kinetic terms, and consequently the correct Poisson bracket structure, it is necessary to obtain the second order Lagrangian without referring to the background equations of motion.

\section{Conclusions}

The main result of this paper was the construction of the Hamiltonian [Eq.~\eqref{ham27}] describing the dynamics of linear cosmological perturbations on a homogeneous and isotropic background with arbitrary curved spatial sections filled with a general fluid with entropy, with the zeroth and second order Hamiltonian terms given by Eqs.~\eqref{def:dHA2ss} and \eqref{ham270}, respectively. It is a rather simple Hamiltonian obtained without ever using the background equations of motion, which can be immediately used for the canonical quantization of the perturbations and background, as it was done in Refs.~\cite{Peter2005,Pinho2007,Falciano2009} for the case of hydrodynamical perfect fluids and scalar fields without a potential.

Similar forms of the Hamiltonian $\dHA^{(2,s)}$, obtained in the case of matter described by one scalar field, have been derived in the literature and considered in the cosmological context of a one-bubble inflationary universe and in K-inflation, see Refs.~\cite{Garriga1998,Garriga1999}. Besides the generality of our approach, note that our procedure follows the logical steps of going from a Lagrangian system to a Hamiltonian formulation, and the commonly known variables present in the literature, such as $\MS$ as defined in Eq.~\eqref{def:MS}, appear naturally during the process.

Our next steps are, of course, to perform the canonical quantization of the model, and use the same Faddeev-Jackiw reduction method in order to obtain a simplified Hamiltonian for the case of scalar fields with a potential filling a Friedmann model with arbitrary curved spatial sections without ever using the background equations of motions, hence completing the program initiated with Refs.~\cite{Peter2005,Pinho2007,Falciano2009} for the case of linear cosmological perturbations in homogeneous and isotropic backgrounds. Then we can begin the much more involved program of obtaining the Hamiltonians describing the dynamics of linear perturbations in Bianchi models and of second order perturbations in Friedmann models.

\section*{ACKNOWLEDGMENTS}

We would like to thank CNPq of Brazil for financial support.  We also would like to thank ``Pequeno Semin\'{a}rio'' of CBPF's Cosmology Group for useful discussions, comments and suggestions.

\appendix

\section{Kinematic Perturbations}
\label{app:kpert}

In this Appendix we shall obtain the perturbations of the kinematic parameters defining an arbitrary spatial slicing. The space-time foliation can be defined through the normal vector field $\n^\mu$ which in turn can be characterized by its kinematic parameters. We perform these calculations for an arbitrary background with another choice of spatial hyper-surfaces ($\bn^\mu$) with the assumption that both spatial sectioning are global and that the background foliation is geodesic, i.e., $\bn^\mu\cd_\mu\bn^\nu = 0$. However, we are interested in a description in which these sectioning are close in the sense that $\dn^\mu \equiv \n^\mu - \bn^\mu$ has the same order of magnitude as $\ddg_{\mu\nu}$. Thus, at first order we have
\begin{align}\label{eq:norm:n}
\n_\mu\n_\nu\g^{\mu\nu} &\approx -1 - \ddg_{\bn\bn}  + 2\dn_{\bn} = -1, \nonumber\\
& \Rightarrow \dn_{\bn} = \A \equiv \frac{\ddg_{\bn\bn}}{2},
\end{align}
where we have defined the time-time projection of $\ddg_{\mu\nu}$. Defining the spatial projection of $\dn_\mu$, $\sn_\mu = \bhp{\dn_\mu}$ we obtain
\begin{equation}\label{eq:def:dn:1}
\dn_\mu = -\A\bn_\mu + \sn_\mu,
\end{equation}
and for $\dn^\mu \equiv \n^\mu - \bn^\mu$,
\begin{equation}\label{eq:def:dn:2}
\dn^\mu = \A\bn^\mu + \sn^\mu + \B^\mu, \quad \B^\mu \equiv -\bhp{\ddg_{\bn}{}^{\mu}}.
\end{equation}
Finally, the perturbation in the projector $\h_\mu{}^\nu$ is
\begin{equation}
\dhh_{\mu}{}^\nu \equiv \h_\mu{}^\nu - \bh_\mu{}^\nu = \sn_\mu\bn^\nu + \bn_\mu[\sn^\nu + \B^\nu].
\end{equation}

\subsection{Acceleration Field}

Using the expressions above we obtain the perturbation on the acceleration ($\dac_\mu \equiv \ac_\mu - \bac_\mu$) as
\begin{equation}\label{eq:dac:1}
\dac_\mu = \B^\alpha\bEC_{\mu\alpha} + \dot{\sn}_{\mu} -\dot{\A}\bn_\mu + \LL_{\mu\bn}{}^{\bn}.
\end{equation}
From Eq.~\eqref{eq:LL:1} we note that $\LL_{\alpha(\mu\nu)} = -\ddg_{\mu\nu;\alpha} / 2$, hence,
\begin{equation}
\LL_{\mu\bn}{}^{\bn} = -\A_{\scp\mu} + \dot{\A}\bn_\mu - \B_\alpha\bEC^\alpha{}_\mu.
\end{equation}
In terms of the metric perturbation, the perturbation of the acceleration field reads
\begin{equation}
\dac_\mu = \dot{\sn}_{\mu}  -\A_{\scp\mu}.
\end{equation}
The covariant field is expressed by
\begin{align} \nonumber
\dac^\mu &= \dot{\A}\bn^\mu + [\dot{\B}_\alpha+\dot{\sn}_\alpha]\bh^{\alpha\mu} - \LL_{\bn\bn}{}^\mu, \\ \label{eq:dac:2}
\dac^\mu &= \dot{\sn}_{\alpha}\bg^{\alpha\mu} - \A^{\scp\mu},
\end{align}
where the spatial projection of $\ddg_{\mu\nu}$ is defined as
\begin{equation}
\C_{\mu\nu} \equiv \frac{\bhp{\ddg_{\mu\nu}}}{2},
\end{equation}
and the projection $\LL_{\bn\bn}{}_\mu$ can be written as
\begin{equation}
\LL_{\bn\bn}{}_\mu = \dot{\A}\bn_\mu + \A_{\scp\mu} + \dot{\B}_\mu.
\end{equation}

Note also that the global slicing condition $\cd_{[\mu}\n_{\nu]} = \ac_{[\mu}\n_{\nu]}$, reduces at first order to
\begin{equation}
\sn_{[\nu\scp\mu]} = 0.
\end{equation}

\subsection{Extrinsic Curvature}

From its definition $\EC_{\mu\nu} \equiv \hp{\cd_\mu\n_\nu}$, we obtain
\begin{equation}\label{eq:dEC:LLbv:1}
\begin{split}
\dEC_\bn{}^\bn &= 0,\\
\dEC_\bn{}^\nu &= -[\sn^\sigma+\B^\sigma]\bEC_\sigma{}^\nu, \\
\dEC_\mu{}^\bn &= -\sn_\sigma\bEC_\mu{}^\sigma, \\
\bhp{\dEC_\mu{}^\nu} &= \A\bEC_\mu{}^\nu + [\sn^\nu+\B^\nu]_{\scp\mu} - \bhp{\LL_{\bn\mu}{}^\nu}.
\end{split}
\end{equation}
The projection $\bhp{\LL_{\bn\mu}{}^\nu}$ can be calculated directly from Eq.~\eqref{eq:LL:1}
\begin{equation}\label{eq:LL:aux1}
\bhp{\LL_{\bn\mu}{}^\nu} = -\left[\frac{\B_\mu{}^{\scp\nu} - \B^\nu{}_{\scp\mu}}{2} + \bh_{\mu\alpha}\bh^{\nu\beta}\dot{\C}_\beta{}^\alpha \right].
\end{equation}
Substituting Eq.~\eqref{eq:LL:aux1} back in Eq.~\eqref{eq:dEC:LLbv:1} we have
\begin{equation}\label{eq:dEC}
\bhp{\dEC_\mu{}^\nu} = \A\bEC_\mu{}^\nu + \sn^\nu{}_{\scp\mu} + \frac{\B_\mu{}^{\scp\nu} + \B^\nu{}_{\scp\mu}}{2} + \bh_{\mu\alpha}\bh^{\nu\beta}\dot{\C}_\beta{}^\alpha.
\end{equation}

Equation~\eqref{eq:dEC:LLbv:1}, relates $\bhp{\LL_{\mu\bn}{}^\beta}$ with the extrinsic curvature perturbation $\dEC_\mu{}^\beta$. To obtain an expression for $\bhp{\LL_{\mu\nu}{}^\bn}$ we have
\begin{equation}\label{eq:dEC:LLbv:2}
\begin{split}
\dEC_{\bn\bn} &= 0, \\
\dEC_{\mu\bn} &= -(\sn^\sigma+\B^\sigma)\bEC_{\sigma\mu}, \\
\bhp{\dEC_{\mu\nu}} &= -\A\bEC_{\mu\nu} + \sn_{\nu\scp\mu} + \bhp{\LL_{\mu\nu}{}^\bn}.
\end{split}
\end{equation}

\subsection{Expansion Factor}

Recalling that $\EX\equiv\EC_{\mu}{}^\mu$, Eq.~\eqref{eq:dEC} gives us
\begin{equation}\label{eq:dEX}
\dEX = \A\bEX + [\sn^\mu+\B^\mu]{}_{\scp\mu} + \dot{\C}.
\end{equation}

\subsection{Spatial Curvature}

We can rewrite Eq.~\eqref{eq:def:SR} by substituting the spatial derivative with its explicit form in terms of the covariant derivative. In this manner, we obtain
\begin{equation}
\SR_{\mu\nu\alpha}{}^\beta = 2\EC_{[\mu}{}^\beta\EC_{\nu]\alpha} + \hp{\R_{\mu\nu\alpha}{}^\beta}.
\end{equation}
Therefore, the perturbation at first order becomes
\begin{equation*}
\begin{split}
\dSR_{\mu\nu\alpha}{}^\beta &= 2\dEC_{[\mu}{}^\beta\bEC_{\nu]\alpha} + 2\bEC_{[\mu}{}^\beta\dEC_{\nu]\alpha}\\
&+ \delta\hp{\bR_{\mu\nu\alpha}{}^\beta} + \bhp{\dR_{\mu\nu\alpha}{}^\beta}.
\end{split}
\end{equation*}
Using Eq.~\eqref{eq:pR:R}, the last term above can be expressed as
\begin{align*}
\frac12&\bhp{\dR_{\mu\nu\alpha}{}^\beta} = \bhp{\LL_{\alpha[\nu}{}^\beta{}_{;\mu]}}= \\
&\bhp{\LL_{\alpha[\nu}{}^\beta}{}_{\scp\mu]} - \bhp{\LL_{\bn[\nu}{}^\beta}\bEC_{\mu]\alpha} - \bEC_{[\mu}{}^\beta\bhp{\LL_{\nu]\alpha}{}^\bn}.
\end{align*}

We can use Eqs.~\eqref{eq:dEC:LLbv:1} and \eqref{eq:dEC:LLbv:2} to express the tensor $\LL_{\mu\nu}{}^\beta$ in terms of the extrinsic curvature. In this way, the time projections of the perturbed spatial curvature  are
\begin{equation}\label{eq:dSRa:LL:1}
\begin{split}
\dSR_{\bn\nu\alpha}{}^\beta &= -(\sn^\sigma+\B^\sigma)\bSR_{\sigma\nu\alpha}{}^\beta, \\
\dSR_{\mu\bn\alpha}{}^\beta &= -(\sn^\sigma+\B^\sigma)\bSR_{\mu\sigma\alpha}{}^\beta, \\
\dSR_{\mu\nu\bn}{}^\beta &= -(\sn^\sigma+\B^\sigma)\bSR_{\mu\nu\sigma}{}^\beta, \\
\dSR_{\mu\nu\alpha}{}^\bn &= -\sn_\sigma\bSR_{\mu\nu\alpha}{}^\sigma, \\
\end{split}
\end{equation}
while its spatial projection can be rewritten as
\begin{equation}\label{eq:dSRb:LL:1}
\begin{split}
\bhp{\dSR_{\mu\nu\alpha}{}^\beta} &= 4\left[\sn_{[\nu}\bEC_{\mu][\sigma\scp\alpha]} + \left(\sn_{[\sigma}\bEC_{\alpha][\nu}\right){}_{\scp\mu]}\right]\bh^{\sigma\beta} \\
&+2\left(\B^\beta\bEC_{\alpha[\nu}\right){}_{\scp\mu]} + 2\bhp{\LL_{\alpha[\nu}{}^\beta}{}_{\scp\mu]}.
\end{split}
\end{equation}
In addition, by projecting Eq.~\eqref{eq:LL:1} we have
\begin{equation}\label{eq:SLL}
\bhp{\LL_{\alpha\nu}{}^\beta} = \SLL_{\alpha\nu}{}^\beta - \B^\beta\bEC_{\alpha\nu},
\end{equation}
where we have defined the tensor
\begin{equation}
\SLL_{\alpha\nu}{}^\beta \equiv -{\gamma^{\sigma\beta}}\left[\C_{\alpha\sigma\scp\nu} + \C_\sigma{}_{\nu\scp\alpha} - \C_{\alpha\nu\scp\sigma}\right],
\end{equation}
and its contractions
\begin{align}
\SLLa_\alpha &\equiv \SLL_{\alpha\nu}{}^\nu = -{\C_{\scp\alpha}}, \\
\SLLb^\beta &\equiv \SLL_{\alpha\nu}{}^\beta\bh^{\alpha\nu} = -2\C_\mu{}^{\sigma}{}_{\scp\sigma} + \C_{\scp\mu}.
\end{align}

In terms of the tensor $\SLL_{\alpha\nu}{}^\beta$, the projection of the spatial Riemann tensor becomes
\begin{equation}\label{eq:dSRb:LL:2}
\begin{split}
\bhp{\dSR_{\mu\nu\alpha}{}^\beta} &= 4\left[\sn_{[\nu}\bEC_{\mu][\sigma\scp\alpha]} + \left(\sn_{[\sigma}\bEC_{\alpha][\nu}\right){}_{\scp\mu]}\right]\bh^{\sigma\beta} \\
& + 2{\SLL_{\alpha[\nu}{}^\beta}{}_{\scp\mu]}.
\end{split}
\end{equation}

Alternatively, we can also express the perturbation of the spatial Riemann tensor in terms of metric perturbations, i.e.
\begin{equation}\label{eq:dSRb:LL:3}
\begin{split}
\bhp{\dSR_{\mu\nu\alpha}{}^\beta} &= 4\left[\sn_{[\nu}\bEC_{\mu][\sigma\scp\alpha]} + \left(\sn_{[\sigma}\bEC_{\alpha][\nu}\right){}_{\scp\mu]}\right]\bh^{\sigma\beta} \\
&+2\left(\C_{\nu[\alpha\scp\sigma]\mu} - \C_{\mu[\alpha\scp\sigma]\nu}\right)\gamma^{\sigma\beta} \\
&-\left(\bSR_{\mu\nu\alpha}{}^\lambda\C_{\sigma\lambda} + \bSR_{\mu\nu\sigma}{}^\lambda\C_{\lambda\alpha}\right)\gamma^{\sigma\beta}.
\end{split}
\end{equation}

\subsubsection*{Spatial Ricci Tensor}

We can obtain the normal projection of the perturbation on the spatial Ricci tensor by contracting the indexes $\nu$ and $\beta$ in Eqs.~\eqref{eq:dSRa:LL:1} and Eq.~\eqref{eq:dSRb:LL:2}, which gives
\begin{equation}
\dSR_{\bn\alpha} = -(\sn^\sigma+\B^\sigma)\bSR_{\sigma\alpha}.
\end{equation}
Similarly, its spatial projection is
\begin{equation}\label{eq:dSR:ricci:1}
\begin{split}
\bhp{\dSR_{\mu\alpha}} &= 2\bigg(\sn^\nu\bEC_{\nu(\mu\scp\alpha)} + \sn_{(\mu}\bEC_{\alpha)}{}^\nu{}_{\scp\nu} -\sn_{(\mu}\bEX_{\scp\alpha)} \\
& + \sn^\nu\bEC_{\mu\alpha\scp\nu} +\sn_{\nu\scp(\mu}\bEC^\nu{}_{\alpha)} \bigg) - \sn^\nu{}_{\scp\nu}\bEC_{\mu\alpha} \\
&- \sn_{\alpha\scp\mu}\bEX +\SLLa_{\alpha\scp\mu} - {\SLL_{\alpha\mu}{}^\nu}{}_{\scp\nu}.
\end{split}
\end{equation}
Using the spatial projection given in Eq.~\eqref{eq:dSRb:LL:3}, one has
\begin{equation}\label{eq:dSR:ricci:2}
\begin{split}
&\bhp{\dSR_{\mu\alpha}} = 2\bigg(\sn^\nu\bEC_{\nu(\mu\scp\alpha)} + \sn_{(\mu}\bEC_{\alpha)}{}^\nu{}_{\scp\nu} -\sn_{(\mu}\bEX_{\scp\alpha)} \\
&+ \sn^\nu\bEC_{\mu\alpha\scp\nu} + \sn_{\nu\scp(\mu}\bEC^\nu{}_{\alpha)} \bigg) - \sn^\nu{}_{\scp\nu}\bEC_{\mu\alpha} - \sn_{\alpha\scp\mu}\bEX \\
&+2\C^\nu{}_{(\alpha\scp\mu)\nu} - \C_{\scp\alpha\mu} - \C_{\alpha\mu\scp\nu}{}^{\scp\nu}.
\end{split}
\end{equation}

\subsubsection*{Spatial Curvature Scalar}

The spatial curvature scalar is given by $\SR = \SR_{\mu\alpha}\g^{\mu\alpha}$. Therefore, its perturbation is given by $\dSR = \dSR_{\mu\alpha}\bg^{\mu\alpha} - \bSR_{\mu\alpha}\ddg^{\mu\alpha}$. Recalling Eq.~\eqref{eq:dSR:ricci:1} we obtain
\begin{equation}\label{eq:dSR:scalar:1}
\begin{split}
\dSR &= 4\sn^\nu\bEC_{\nu}{}^\alpha{}_{\scp\alpha} + 2\sn_{\nu\scp\mu}\bEC^{\nu\mu} - 2\sn^\nu{}_{\scp\nu}\bEX \\
&+ \SLLa^\mu{}_{\scp\mu} - {\SLLb^\mu}{}_{\scp\mu} - 2\bSR_{\mu\nu}\C^{\mu\nu}.
\end{split}
\end{equation}
Alternatively, from Eq.~\eqref{eq:dSR:ricci:2} we have
\begin{equation}\label{eq:dSR:scalar:2}
\begin{split}
\dSR &= 4\sn^\nu\bEC_{\nu}{}^\alpha{}_{\scp\alpha} + 2\sn_{\nu\scp\mu}\bEC^{\nu\mu} - 2\sn^\nu{}_{\scp\nu}\bEX \\
&+ 2\C^{\mu\nu}{}_{\scp\mu\nu}-2\C_{\scp\nu}{}^{\scp\nu}-2\bSR_{\mu\nu}\C^{\mu\nu}.
\end{split}
\end{equation}

\section{Gravitational Perturbation Lagrangian}
\label{sec:pert:L2}

In this section we shall rewrite the gravitational Lagrangian in terms of the perturbations on the kinetic tensors obtained in Appendix~\eqref{app:kpert}. These perturbations were obtained by considering two different spatial sectioning. However, it is more appealing from a physical point of view, to construct the perturbed quantities with respect to the same spatial sections.

We can achieve this change in the description of the perturbations by imposing that both sectioning have the same normal one-form field $\bn_\mu$. Note however that their normalization shall not be equal, since each of them is normalized with its appropriated metric. From Eqs.~\eqref{eq:norm:n} and \eqref{eq:def:dn:1}, we see that in this case the field $\sn_\mu$ vanishes. Therefore, we can obtain all perturbations for the case where both slicing have the same normal field by setting $\sn_\mu$ equal to zero. This kind of kinetic variables are described in details in~\cite{Kodama1984}.

First, we shall develop the kinetic part of the second order gravitational Lagrangian Eq.~\eqref{eq:def:dLAk2}. This term can be decomposed as
$$\LL_{\mu\nu}{}^\gamma\LL_\gamma{}^{(\mu\nu)} - \LLa_\mu\LLb^\mu = \sum_{n=1}^{5} l_n,$$
where
\begin{align}
&l_1 = -\left(\bhp{\LL_{\bn\nu}{}^\gamma}\bhp{\LL_\gamma{}^{\bn\nu}} + 2\bhp{\LL_{\bn\mu}{}^\gamma}\bhp{\LL_\gamma{}^{\mu\bn}}\right), \\ \label{eq:def:l2}
&l_2 = 2\bhp{\LL_{\bn\bn}{}^\gamma}\bhp{\LL_{\gamma\bn}{}^\bn} + \bhp{\LL_{\gamma\bn}{}^\bn}\bhp{\LL^\gamma{}_{\bn}{}^\bn}, \\
&l_3 = -\bhp{\LLa_\mu}\bhp{\LLb^\mu}, \\
&l_4 = \LLa_\bn\LLb^\bn - \LL_{\bn\bn}{}^\bn\LL_{\bn\bn}{}^\bn,\\
&l_5 = \bhp{\LL_{\mu\nu}{}^\gamma}\bhp{\LL_\gamma{}^{\mu\nu}}.
\end{align}
Using Eqs.~\eqref{eq:dEC:LLbv:1} and \eqref{eq:dEC:LLbv:2} we obtain
\begin{align*}
l_1 &= -(\A\bEC_{\mu}{}^\gamma+\B^\gamma{}_{\scp\mu})(\A\bEC_{\gamma}{}^\mu+\B^\mu{}_{\scp\gamma}) + \dEC_\mu{}^\gamma\dEC_\gamma{}^\mu \nonumber\\
&-2(\A\bEC_{\mu}{}^\gamma+\B^\gamma{}_{\scp\mu})(\A\bEC_\gamma{}^\mu + 2\bEC_\gamma{}^\sigma\C_\sigma{}^\mu) \nonumber\\
&+2(\A\bEC_\mu{}^\gamma + 2\C_\mu{}^\sigma\bEC_\sigma{}^\gamma)\dEC_\gamma{}^\mu.
\end{align*}
In order to obtain a Lagrangian that is quadratic in the terms involving time derivative, we can use  Eq.~\eqref{eq:dEC} to rewrite the quantity $\C_\mu{}^\sigma\bEC_\sigma{}^\gamma\dEC_\gamma{}^\mu$ modifying $l_1$ to
\begin{align*}
l_1 &= -(\A\bEC_{\mu}{}^\gamma+\B^\gamma{}_{\scp\mu})(3\A\bEC_{\gamma}{}^\mu+\B^\mu{}_{\scp\gamma}) +\dEC_\mu{}^\gamma\dEC_\gamma{}^\mu \\
&+ 4\C^{\mu\sigma}\bEC_\sigma{}^\gamma\B_{[\mu\scp\gamma]} +2\A\bEC_\mu{}^\gamma\dEC_\gamma{}^\mu+ 4\bh^{\sigma\beta}\bEC_\lambda{}_\alpha\C_\sigma{}^\lambda\dot{\C}_\beta{}^\alpha.
\end{align*}
Additionally, we can rewrite the term $4\bh^{\sigma\beta}\bEC_\lambda{}_\alpha\C_\sigma{}^\lambda\dot{\C}_\beta{}^\alpha$ as
\begin{equation}
\begin{split}
&4\bh^{\sigma\beta}\bEC_\lambda{}_\alpha\C_\sigma{}^\lambda\dot{\C}_\beta{}^\alpha = 2 \bh^{\sigma\beta}\bEC_\lambda{}_\alpha\lie_\bn\left[\C_\sigma{}^\lambda{\C}_\beta{}^\alpha\right]=\\
&=\frac{2\lie_\bn\left[\sqrt{-\bg} \bh^{\sigma\beta} \bEC_\lambda{}_\alpha \C_\sigma{}^\lambda \C_\beta{}^\alpha\right]}{\sqrt{-\bg}} \\
&+2\C_\sigma{}^\lambda \C_\beta{}^\alpha\left[2\bEC^{\sigma\beta} \bEC_\lambda{}_\alpha-\bh^{\sigma\beta}(\bEX\bEC_{\lambda\alpha}+\dot{\bEC}_{\lambda\alpha}) \right].
\end{split}
\end{equation}
Therefore, ignoring the surface term, one has
\begin{equation}
\begin{split}
l_1 &= -[\A\bEC_{\mu}{}^\gamma+\B^\gamma{}_{\scp\mu}][3\A\bEC_{\gamma}{}^\mu+\B^\mu{}_{\scp\gamma}] \\
&+ \dEC_\mu{}^\gamma\dEC_\gamma{}^\mu +4\C^{\mu\sigma}\bEC_\sigma{}^\gamma\B_{[\mu\scp\gamma]} +2\A\bEC_\mu{}^\gamma\dEC_\gamma{}^\mu \\
&+2\C_\sigma{}^\lambda \C_\beta{}^\alpha\left[2\bEC^{\sigma\beta} \bEC_\lambda{}_\alpha-\bh^{\sigma\beta}[\bEX\bEC_{\lambda\alpha}+\dot{\bEC}_{\lambda\alpha}] \right].
\end{split}
\end{equation}

The second term $l_2$ defined in Eq.~\eqref{eq:def:l2} can be simplified by using Eqs.~\eqref{eq:dac:1} and \eqref{eq:dac:2}. Thus,
\begin{equation*}
l_2 = -\dac_\gamma\dac^\gamma + 2\dot{\B}_\gamma\dac^\gamma - 2\dot{\B}_\gamma\B_\sigma\bEC^{\gamma\sigma} + \B_\gamma\B_\sigma\bEC^{\gamma\lambda}\bEC_\lambda{}^\sigma.
\end{equation*}
Again we shall rewrite the terms involving only one time derivative such that
\begin{equation}
\begin{split}
&2\dot{\B}_\gamma\B_\sigma\bEC^{\gamma\sigma} = \frac{\lie_\bn[\sqrt{-\bg}{\B}_\gamma\B_\sigma\bEC^{\gamma\sigma}]}{\sqrt{-\bg}} \\
&- \B^\alpha\B^\beta(\bEX\bEC_{\alpha\beta} + \dot{\bEC}_{\alpha\beta}) + 4\B_\gamma\B_\sigma\bEC^{\gamma\alpha}\bEC_\alpha{}^\sigma,
\end{split}
\end{equation}
and also
\begin{equation}
\begin{split}
2\dot{\B}_\gamma\dac^\gamma &= -\frac{\lie_\bn[2\sqrt{-\bg}\B_\gamma\A^{\scp\gamma}]}{\sqrt{-\bg}} + 2\left(\B^\gamma\dot{\A}\right)_{\scp\gamma} \\
&-2\bEX\B^\gamma\dac_\gamma - 2\B^\gamma{}_{\scp\gamma}\dot{\A} + 4\B_\gamma\bEC^{\gamma\alpha}\dac_\alpha.
\end{split}
\end{equation}
Accordingly, discarding the surface term, we have
\begin{equation*}
\begin{split}
l_2 &= -\dac_\gamma\dac^\gamma + \B^\alpha\B^\beta[\bEX\bEC_{\alpha\beta} + \dot{\bEC}_{\alpha\beta}] - 3\B_\gamma\B_\sigma\bEC^{\gamma\alpha}\bEC_\alpha{}^\sigma \\
&-2\bEX\B^\gamma\dac_\gamma - 2\B^\gamma{}_{\scp\gamma}\dot{\A} + 4\B_\gamma\bEC^{\gamma\alpha}\dac_\alpha.
\end{split}
\end{equation*}

The next term is $l_3$, which can be written as
\begin{equation}
\begin{split}
l_3 &= -\SLLa_\mu\SLLb^\mu + \dac_\mu(\SLLb^\mu-\SLLa^\mu) + \dac_\mu\dac^\mu \\
&+ \left(\SLLa^\alpha-\bhp{\dac^\alpha}\right)\left(\bEX\B_\alpha + \dot{\B}_\alpha\right),
\end{split}
\end{equation}
where we have used Eqs.~\eqref{eq:dac:1}, \eqref{eq:dac:2} and \eqref{eq:SLL}. The last term above can be re-arranged as
\begin{equation}
\begin{split}
&\left(\SLLa^\alpha-\bhp{\dac^\alpha}\right)\left(\bEX\B_\alpha + \dot{\B}_\alpha\right) = \left[(\dot{\C}-\dot{\A})\B^\alpha\right]_{\scp\alpha}\\
&+\frac{\lie_\bn\left[\sqrt{-\bg}\left(\SLLa^\alpha-\bhp{\dac^\alpha}\right)\B_\alpha\right]}{\sqrt{-\bg}}  \\
&-(\dot{\C}-\dot{\A})\B^\mu{}_{\scp\mu} + 2\bEC^{\alpha\beta}(\SLLa_\beta-\bhp{\dac_\beta})\B_\alpha.
\end{split}
\end{equation}\linebreak
Hence, ignoring the total derivative terms, we obtain
\begin{equation}
\begin{split}
l_3 &= -\SLLa_\mu\SLLb^\mu + \dac_\mu(\SLLb^\mu-\SLLa^\mu) + \dac_\mu\dac^\mu \\
&-(\dot{\C}-\dot{\A})\B^\mu{}_{\scp\mu} + 2\bEC^{\alpha\beta}(\SLLa_\beta-\bhp{\dac_\beta})\B_\alpha.
\end{split}
\end{equation}

The term $l_4$, using Eqs.~\eqref{eq:dEC:LLbv:1}, \eqref{eq:dEC:LLbv:2} and \eqref{eq:dEX}, reads
\begin{equation}
\begin{split}
l_4 &= \B^\mu{}_{\scp\mu}\left(2\A\bEX + \B^\nu{}_{\scp\nu}+\dot{\C}+\dot{\A}\right) \\
&+\A^2\bEX^2-\dEX^2+2\A\dot{\A}\bEX+2\bEC^{\sigma\gamma}\C_{\sigma\gamma}\left(\dot{\A}-\dot{\C}\right).
\end{split}
\end{equation}
Once more, we rewrite
\begin{align*}
&\A^2\bEX^2+2\A\dot{\A}\bEX = \frac{\lie_\bn[\sqrt{-\bg}\A^2\bEX]}{\sqrt{-\bg}}-\dot{\bEX}\A^2, \\
&2\bEC^{\sigma\gamma}\C_{\sigma\gamma}\dot{\A} = \frac{2\lie_\bn[\sqrt{-\bg}\bEC_\alpha{}^\beta\C_\beta{}^\alpha\A]}{\sqrt{-\bg}} - 2\A\bEC_\alpha{}^\beta\dEC_\beta{}^\alpha\\
&-2(\bEX\bEC_\alpha{}^\beta+\dot{\bEC}_\alpha{}^\beta)\C_\beta{}^\alpha\A +2\A^2\bEC_\alpha{}^\beta\bEC_\beta{}^\alpha+2\A\bEC_\alpha{}^\beta\B^\alpha{}_{\scp\beta}.
\end{align*}

Then, ignoring the total derivatives, $l_4$ becomes
\begin{equation}
\begin{split}
l_4 &= \B^\mu{}_{\scp\mu}\left(2\A\bEX + \B^\nu{}_{\scp\nu}+\dot{\C}+\dot{\A}\right) - \dEX^2 \\
&-\dot{\bEX}\A^2 - 2\bEC^{\sigma\gamma}\C_{\sigma\gamma}\dot{\C}-2(\bEX\bEC_\alpha{}^\beta+\dot{\bEC}_\alpha{}^\beta)\C_\beta{}^\alpha\A \\
& - 2 \A\bEC_\alpha{}^\beta\dEC_\beta{}^\alpha + 2\A^2\bEC_\alpha{}^\beta\bEC_\beta{}^\alpha + 2\A\bEC_\alpha{}^\beta\B^\alpha{}_{\scp\beta}.
\end{split}
\end{equation}

The last term $l_5$ is simply
\begin{equation}
l_5 = \SLL_{\mu\nu}{}^\gamma\SLL_\gamma{}^{\mu\nu} - 2\SLL_\gamma{}^{\mu\nu}\B^\gamma\bEC_{\mu\nu} + \B^\gamma\B^\sigma\bEC_{\gamma\alpha}\bEC^\alpha{}_\sigma,
\end{equation}
where we have used Eq.~\eqref{eq:LL:1} to rewrite $\LL_{\mu\nu}{}^\gamma$ in terms of $\SLL_{\mu\nu}{}^\gamma$.
\begin{widetext}
Collecting all these five terms we obtain
\begin{equation}\label{eq:LL_LL:aux}
\begin{split}
\sum_{n=1}^{5} l_n &= \dEC_\mu{}^\nu\dEC_\nu{}^\mu - \dEX^2 -[\A\bEC_{\mu}{}^\gamma+\B^\gamma{}_{\scp\mu}][\A\bEC_{\gamma}{}^\mu+\B^\mu{}_{\scp\gamma}] + 4\C^{\mu\sigma}\bEC_\sigma{}^\gamma\B_{[\mu\scp\gamma]}\\
&+2\C_\sigma{}^\lambda \C_\beta{}^\alpha\left[2\bEC^{\sigma\beta} \bEC_\lambda{}_\alpha-\bh^{\sigma\beta}(\bEX\bEC_{\lambda\alpha}+\dot{\bEC}_{\lambda\alpha}) \right] +\B^\alpha\B^\beta[\bEX\bEC_{\alpha\beta} + \dot{\bEC}_{\alpha\beta}]- 3\B_\gamma\B_\sigma\bEC^{\gamma\alpha}\bEC_\alpha{}^\sigma \\
&-2\bEX\B^\gamma\dac_\gamma + 2\B_\gamma\bEC^{\gamma\alpha}\dac_\alpha -\SLLa_\mu\SLLb^\mu + \dac_\mu[\SLLb^\mu-\SLLa^\mu] + 2\bEC^{\alpha\beta}\SLLa_\beta\B_\alpha +\B^\mu{}_{\scp\mu}\left[2\A\bEX + \B^\nu{}_{\scp\nu}\right] \\
&-\dot{\bEX}\A^2 - 2\bEC^{\sigma\gamma}\C_{\sigma\gamma}\dot{\C} - 2[\bEX\bEC_\alpha{}^\beta+\dot{\bEC}_\alpha{}^\beta]\C_\beta{}^\alpha\A + \SLL_{\mu\nu}{}^\gamma\SLL_\gamma{}^{\mu\nu} - 2\SLL_\gamma{}^{\mu\nu}\B^\gamma\bEC_{\mu\nu} + \B^\gamma\B^\sigma\bEC_{\gamma\alpha}\bEC^\alpha{}_\sigma.
\end{split}
\end{equation}
\end{widetext}
In order to simplify further the above equation, we first note that we have the following identities relating the Ricci tensor and the kinetic variables,
\begin{align}
\dot{\bEX} &= -\bEC_\mu{}^\nu\bEC_\nu{}^\mu - \bR_{\bn\bn}, \\ \label{eq:dotbEC}
\dot{\bEC}_{\mu\nu} + \bEX\bEC_{\mu\nu} &= \bhp{\bR_{\mu\nu}}-\bSR_{\mu\nu} + 2\bEC_\mu{}^\gamma\bEC_{\gamma\nu}, \\
\bhp{\bR_{\mu\bn}} &= \bEC_\mu{}^\nu{}_{\scp\nu} - \bEX_{\scp\mu}.
\end{align}

These are the exactly combinations of background variables appearing in Eq.~\eqref{eq:LL_LL:aux}. In particular, the term quadratic in $\A$ is
\begin{equation}
l_{\A\A} \equiv -(\bEC_\mu{}^\nu\bEC_\nu{}^\mu + \dot{\EX})\A^2 = \bR_{\bn\bn}\A^2.
\end{equation}
Repeating this procedure for the terms quadratic in $\B_\mu$ we first have that
\begin{equation}
\begin{split}
\B^\gamma{}_{\scp\mu}\B^\mu{}_{\scp\gamma} &= \left(\B^\gamma\B^\mu{}_{\scp\gamma}-\B^\mu\B^\gamma{}_{\scp\gamma}\right)_{\scp\mu} \\
&+ (\B^\mu{}_{\scp\mu})^2 - \B^\gamma\B^\sigma\bSR_{\gamma\sigma}.
\end{split}
\end{equation}
Hence, without the total derivatives, we obtain
\begin{equation}
\begin{split}
l_{\B\B} &\equiv \B^\alpha\B^\beta(\bEX\bEC_{\alpha\beta} + \dot{\bEC}_{\alpha\beta})- 2\B_\gamma\B_\sigma\bEC^{\gamma\alpha}\bEC_\alpha{}^\sigma \\
&-\B^\gamma{}_{\scp\mu}\B^\mu{}_{\scp\gamma} + [\B^\mu{}_{\scp\mu}]^2 \\
&=\B^\gamma\B^\sigma\bR_{\gamma\sigma}.
\end{split}
\end{equation}
The cross terms of $\A$ and $\B_\mu$ are
\begin{equation}
\begin{split}
l_{\A\B} &\equiv 2\left[\bEX(\A\B^\gamma)_{\scp\gamma} - \bEC_\mu{}^\gamma(\A\B^\mu)_{\scp\gamma}\right], \\
&= 2\left[\bEX \A\B^\gamma - \bEC_\mu{}^\gamma\A\B^\mu\right]_{\scp\gamma} + 2\bR_{\mu\bn}\B^\mu\A.
\end{split}
\end{equation}

The next possibility is the cross terms between $\A$ and $\C_{\mu\nu}$, i.e.,
\begin{equation}
\begin{split}
l_{\A\C} &\equiv \dac_\mu(\SLLb^\mu - \SLLa^\mu) - 2(\bEX\bEC_\mu{}^\nu+\dot{\bEC}_\mu{}^\nu)\C_\nu{}^\mu\A, \\
&= -\left[\A(\SLLb^\mu - \SLLa^\mu)\right]_{\scp\mu} - \A\dSR - 2\bR_\mu{}^\nu\C_\nu{}^\mu\A,
\end{split}
\end{equation}
where we have used Eq.~\eqref{eq:dSR:scalar:1}. The last cross terms are between $\B_\mu$ and $\C_{\mu\nu}$, given by
\begin{equation*}
l_{\B\C} \equiv 4\C^{\mu\sigma}\bEC_\sigma{}^\gamma\B_{[\mu\scp\gamma]} + 2\left(\C_{\mu\nu}{}_{\scp\gamma}- \C_{\scp\mu}\bh_{\nu\gamma}\right)\bEC^{\mu\nu}\B^\gamma.
\end{equation*}

The extrinsic curvature is given by $3\bEC_{\mu\nu}=\bEX_{\mu\nu} \bh_{\mu\nu}+3\bSH_{\mu\nu}$. However, due to the anti-symmetry in the above $\C$'s terms, only the shear contributes. Thus, we can substitute the extrinsic curvature by the shear in the above equation
\begin{equation*}
l_{\B\C} = 4\C^{\mu\sigma}\bSH_\sigma{}^\gamma\B_{[\mu\scp\gamma]} + 2\left(\C_{\mu\nu}{}_{\scp\gamma}- \C_{\scp\mu}\bh_{\nu\gamma}\right)\bSH^{\mu\nu}\B^\gamma.
\end{equation*}
There are also the terms quadratic in $\C_{\mu\nu}$, i.e.
\begin{equation*}
\begin{split}
l_{\C\C} &\equiv 2\C_\sigma{}^\lambda \C_\beta{}^\alpha\left(2\bEC^{\sigma\beta} \bEC_\lambda{}_\alpha-\bh^{\sigma\beta}[\bEX\bEC_{\lambda\alpha}+\dot{\bEC}_{\lambda\alpha}]\right) \\
&+ \SLL_{\mu\nu}{}^\gamma\SLL_\gamma{}^{\mu\nu}  -\SLLa_\mu\SLLb^\mu- 2\bEC^{\sigma\gamma}\C_{\sigma\gamma}\dot{\C}.
\end{split}
\end{equation*}
The last two terms above may be rewritten as
\begin{align*}
\SLL_{\mu\nu}{}^\gamma\SLL_\gamma{}^{\mu\nu} &= -\left(\SLL_{\mu\nu}{}^\gamma\C^{\mu\nu}\right)_{\scp\gamma} + \C^{\mu\nu}\SLL_{\mu\nu}{}^\gamma{}_{\scp\gamma}, \\
-\SLLa_\mu\SLLb^\mu &= \left[\SLLa_\mu\C^{\mu\sigma}-\C\left[\SLLa^\sigma-\SLLb^\sigma\right]\right]_{\scp\sigma} \\
&+\frac{\C}{2}\left[\SLL_a^\mu-\SLLb^\mu\right]_{\scp\mu}-\SLLa_{\mu\scp\nu}\C^{\mu\nu}.
\end{align*}
Hence, ignoring the surface terms, we have
\begin{equation}
\begin{split}
l_{\C\C} &= 4\C_\sigma{}^\lambda \C_\beta{}^\alpha\left(\bSH^{\sigma\beta} \bSH_\lambda{}_\alpha-\bh^{\sigma\beta}\bSH_\lambda{}^\gamma\bSH_{\gamma\alpha}\right) \\
&-\C_\mu{}^\nu\dSR_\nu{}^\mu + \frac{\C}{2}\dSR -2\C_\sigma{}^\gamma\C_\gamma{}^\alpha\bR_\alpha{}^\sigma \\
&+ \C\C_\mu{}^\nu\bSR_\nu{}^\mu- 2\bEC^{\sigma\gamma}\C_{\sigma\gamma}\dot{\C}.
\end{split}
\end{equation}
It is interesting to perform one last simplification by combining the last two terms to change from the spatial Ricci tensor $\bSR_{\nu\mu}$ to the full Ricci tensor $\bR_{\nu\mu}$.

Using Eq.~\eqref{eq:dotbEC}, we can rewrite the last two terms above as
\begin{equation*}
\begin{split}
&\C\C_\mu{}^\nu\bSR_\nu{}^\mu - 2\bEC^{\sigma\gamma}\C_{\sigma\gamma}\dot{\C}=-\frac{\lie_\bn\left[\sqrt{-\bg}\C\C_\mu{}^\nu\bEC_\nu{}^\mu\right]}{\sqrt{-\bg}} \\
&+ \C\C_\mu{}^\nu\bR_\nu{}^\mu + \C\dot{\C}_\mu{}^\nu\bEC_\nu{}^\mu - \dot{\C}\C_\mu{}^\nu\bEC_\nu{}^\mu.
\end{split}
\end{equation*}
Finally, ignoring the total derivatives and rewriting it in terms of the kinetic perturbations, we obtain
\begin{equation*}
\begin{split}
l_{\C\C} &= \C\C_\mu{}^\nu\bR_\nu{}^\mu + \left(\C\dEC_\mu{}^\nu - \dEX\C_\mu{}^\nu\right)\bSH_\nu{}^\mu \\
&+\left[\left(\bEX\A + \B^\gamma{}_{\scp\gamma}\right)\C_\mu{}^\nu - \left(\A\bEC_\nu{}^\mu + \B_\mu{}^{\scp\nu}\right)\C\right]\bSH_\nu{}^\mu.
\end{split}
\end{equation*}
\begin{widetext}
Putting all the above results together we have that the kinetic term $\dLA_{\g\kin}^{(2)}$ [Eq.~\eqref{eq:def:dLAk2}] is given by
\begin{equation}
\begin{split}
\dLA_{\g\kin}^{(2)} &= \frac{\sqrt{-\bg}}{2\kp}\Bigg\{\dEC_\mu{}^\nu\dEC_\nu{}^\mu - \dEX^2 -\C_\mu{}^\nu\dSR_\nu{}^\mu +\left(\frac{\C}{2}- \A\right)\dSR + 4\C^{\mu\sigma}\bSH_\sigma{}^\gamma\B_{[\mu\scp\gamma]} + 2\left(\C_{\mu\nu}{}_{\scp\gamma}- \C_{\scp\mu}\bh_{\nu\gamma}\right)\bSH^{\mu\nu}\B^\gamma \\
&+ 4\C_\sigma{}^\lambda \C_\beta{}^\alpha\left(\bSH^{\sigma\beta} \bSH_\lambda{}_\alpha-\bh^{\sigma\beta}\bSH_\lambda{}^\gamma\bSH_{\gamma\alpha}\right) + \left[\left(\bEX\A + \B^\gamma{}_{\scp\gamma}-\dEX\right)\C_\mu{}^\nu - \left(\A\bEC_\nu{}^\mu + \B_\mu{}^{\scp\nu}-\dEC_\mu{}^\nu\right)\C\right]\bSH_\nu{}^\mu\\
&+\bR_{\bn\bn}\A^2+\B^\gamma\B^\sigma\bR_{\gamma\sigma} -2\C_\sigma{}^\gamma\C_\gamma{}^\alpha\bR_\alpha{}^\sigma+\C\C_\mu{}^\nu\bR_\nu{}^\mu - 2\bR_\mu{}^\nu\C_\nu{}^\mu\A + 2\bR_{\mu\bn}\B^\mu\A\Bigg\}.
\end{split}
\end{equation}
Using the decomposition of $\ddg_{\mu\nu}$, the potential term $\dLA_{\g\pot}^{(2)}$ [Eq.~\eqref{eq:def:dLAp2}] reads
\begin{equation}
\begin{split}
\dLA_{\g\pot}^{(2)} &= \frac{\sqrt{-\bg}}{2\kp}\Bigg\{\bG_{\bn\bn}\left(\B_\mu\B^\mu-2\A^2-2\A\C\right) +\bG_{\bn\mu}\left[4\B_\alpha\C^{\alpha\mu} - 2(\A+\C)\B^\mu\right] \\
&+\bG_{\mu\nu}\left[4\C^{\mu\alpha}\C_\alpha{}^\nu-\B^\mu\B^\nu + 2\C^{\mu\nu}(\A-\C)\right] +\frac{\bR}{4}\left(4\C_{\mu\nu}\C^{\mu\nu} - 2\B_\mu\B^\mu +2\A^2-2\C^2+4\A\C\right)\Bigg\}.
\end{split}
\end{equation}
Therefore, summing these two terms, we obtain the gravitational second order Lagrangian
\begin{equation}\label{eq:final:dLA2}
\begin{split}
\dLA_{\g}^{(2)} &= \frac{\sqrt{-\bg}}{2\kp}\Bigg\{\dEC_\mu{}^\nu\dEC_\nu{}^\mu - \dEX^2 -\C_\mu{}^\nu\dSR_\nu{}^\mu +\left(\frac{\C}{2}- \A\right)\dSR + 4\C^{\mu\sigma}\bSH_\sigma{}^\gamma\B_{[\mu\scp\gamma]} + 2\left(\C_{\mu\nu}{}_{\scp\gamma}- \C_{\scp\mu}\bh_{\nu\gamma}\right)\bSH^{\mu\nu}\B^\gamma \\
&+ 4\C_\sigma{}^\lambda \C_\beta{}^\alpha\left(\bSH^{\sigma\beta} \bSH_\lambda{}_\alpha-\bh^{\sigma\beta}\bSH_\lambda{}^\gamma\bSH_{\gamma\alpha}\right) + \left[\left(\bEX\A + \B^\gamma{}_{\scp\gamma}-\dEX\right)\C_\mu{}^\nu - \left(\A\bEC_\nu{}^\mu + \B_\mu{}^{\scp\nu}-\dEC_\mu{}^\nu\right)\C\right]\bSH_\nu{}^\mu\\
&+\bG_{\bn\bn}\left(\B_\mu\B^\mu-\A^2-2\A\C\right) +\bG_{\bn\mu}\left(4\B_\alpha\C^{\alpha\mu}- 2\C\B^\mu\right)  + \bG_{\mu\nu}\left(2\C^{\mu\alpha}\C_\alpha{}^\nu - \C^{\mu\nu}\C\right) \Bigg\}.
\end{split}
\end{equation}
\end{widetext}

\section{Matter Perturbation Lagrangian}
\label{sec:pert:Lm2}

In this section we first express the perturbed matter Lagrangian up to second order in terms of the perturbations on the metric and the energy momentum tensor. Then, the spatial splitting is introduced by writing both perturbations in terms of its projections.

Expansion up to second order of the matter action given in Eq.~\eqref{eq:def:ACm} can be written as
\begin{equation}
\AC_\mat[\g_{\mu\nu},\mf_a] = \bAC_\mat + \dAC_\mat^{(1)} + \dAC_\mat^{(2)},
\end{equation}
where $\mf_a$'s represent arbitrary matter fields. We are assuming that the matter Lagrangian does not depend on derivatives of the metric. The first order term is
\begin{equation}
\dLA_\mat^{(1)} = \wbackdec{\frac{\delta\AC_\mat}{\dmf_a}}\dmf_a + \frac{\sqrt{-\bg}}{2}\bT^ {\mu\nu}\ddg_{\mu\nu},
\end{equation}
where the overline means that the quantity should be evaluated with the background fields. To simplify our notation, we define
\begin{equation}
\delta^{\mf}F \equiv \int\dd^4x\wbackdec{\frac{\delta{}F}{\dmf_a}}\dmf_a, \quad \delta^{\g}F \equiv \int\dd^4x\wbackdec{\frac{\delta{}F}{\delta\g_{\mu\nu}}}\dg_{\mu\nu},
\end{equation}
to represent, respectively, the perturbation of $F$ varying only the matter fields or the metric.

The second order part is
\begin{equation}\label{eq:def:ACm2:Appendix}
\begin{split}
\frac{\dLA_\mat^{(2)}}{\sqrt{-\bg}} &=\left[\frac{\delta^{\mf}\T_{\mu\nu}}{2} + \frac{\delta^\g\T_{\mu\nu}}{4}\right]\ddg^{\mu\nu}-
\frac{\bT_{\mu\nu}}{2}\left[\ddg^{\mu\alpha}\ddg_\alpha{}^\nu-\frac{\ddg\ddg^{\mu\nu}}{4}\right] \\
&+\int\dd^4{}y\wbackdec{\frac{\delta^2\AC_\mat}{\dmf_a\dmf_b(y)}}\frac{\dmf_a\dmf_b(y)}{2\sqrt{-\bg}},
\end{split}
\end{equation}
where we have written explicitly only the space-time coordinates that differs from $x$. In terms of the full first order perturbation we have $\dT_\mu{}^\nu = \delta^\g\T_\mu{}^\nu + \delta^\mf\T_\mu{}^\nu$. Thus, the second order matter action can be written as
\begin{equation}
\begin{split}
\frac{\dLA_\mat^{(2)}}{\sqrt{-\bg}} &= \left[\frac{\dT_\mu{}^\nu}{2} - \frac{\delta^\g\T_\mu{}^\nu}{4}\right]\ddg_\nu{}^\mu- \frac{\bT_{\mu\nu}}{4}\left[\ddg^{\mu\alpha}\ddg_\alpha{}^\nu-\frac{\ddg\ddg^{\mu\nu}}{2}\right] \\
&+\int\dd^4{}y\wbackdec{\frac{\delta^2\AC_\mat}{\dmf_a\dmf_b(y)}}\frac{\dmf_a\dmf_b(y)}{2\sqrt{-\bg}}.
\end{split}
\end{equation}

Given an arbitrary foliation defined by the field $\n^\mu$, we define the energy density, energy flux and stress tensor as
\begin{equation}
\ed \equiv \T_{\n\n}, \quad q_\mu \equiv -\hp{\T_{\n\mu}}, \quad \ST_{\mu\nu} \equiv \hp{\T_{\mu\nu}}.
\end{equation}

The perturbation of each projection is related with the components of $\dT_\mu{}^\nu$ by the relations
\begin{align}
\ded &= \dT_\bn{}^\bn - \bq_\mu\B^\mu, \\
\dq_\mu &= -\bhp{\dT_\mu{}^\bn} - \bq_\mu\A + \bn_\mu\B^\gamma\bq_\gamma, \\
\dq^\nu &= -\bhp{\dT_\bn{}^\nu} + \bq^\nu\A - \bST_\mu{}^\nu\B^\mu - \B^\nu\bed, \\
\dST_\mu{}^\nu &= \bhp{\dT_\mu{}^\nu} + \bn_\mu\bST_\alpha{}^\nu\B^\alpha - \bq_\mu\B^\nu,
\end{align}
where $\bn^\mu$ is the background foliation vector field.

Using the above expressions we rewrite the following parts of the Lagrangian in terms of the projections of both metric and energy momentum tensor perturbations. For the energy momentum tensor perturbed with respect to all perturbations one has
\begin{equation*}
\begin{split}
\frac{\dT_\mu{}^\nu\ddg_\nu{}^\mu}{2} &= \A\ded - \B^\mu\dq_\mu + \C_\nu{}^\mu\dST_\mu{}^\nu - \frac{\bed}{2}\B_\mu\B^\mu \\
&+ \bq_\mu(\A\B^\mu + 2\B^\alpha\C_\alpha{}^\mu) - \bST_\mu{}^\nu\frac{\B_\nu\B^\mu}{2}.
\end{split}
\end{equation*}
For the energy momentum tensor perturbed with respect to only metric perturbation
\begin{equation*}
\begin{split}
\frac{\delta^\g\T_\mu{}^\nu\ddg_\nu{}^\mu}{4} &= \frac{\A\delta^\g\ed}{2} - \frac{\B^\mu\delta^\g{}q_\mu}{2} +\frac{\C_\nu{}^\mu}{2}\delta^\g\ST_\mu{}^\nu - \frac{\bed}{4}\B_\mu\B^\mu \\
&+\frac{\bq_\mu}{2}[\A\B^\mu + 2\B^\alpha\C_\alpha{}^\mu] - \frac{\bST_\mu{}^\nu}{2}\frac{\B_\nu\B^\mu}{2}.
\end{split}
\end{equation*}

Concluding, the second order matter Lagrangian for a general combination of scalar fields expressed in terms of the projections of the metric and energy momentum perturbations is given by
\begin{widetext}
\begin{equation}\label{eq:def:ACm2:1}
\begin{split}
&\frac{\dLA_\mat^{(2)}}{\sqrt{-\bg}} = \int\dd^4{}y\wbackdec{\frac{\delta^2\AC_\mat}{\dmf_a\dmf_b(y)}}\frac{\dmf_a\dmf_b(y)}{2\sqrt{-\bg}} - \B^\mu\left(\dq_\mu-\frac{\delta^\g{}q_\mu+\bq_\mu\C}{2}\right) + \C_\nu{}^\mu\left(\dST_\mu{}^\nu-\frac{\delta^\g\ST_\mu{}^\nu+\bST_\mu{}^\nu\A}{2}\right)\\
&+\A\left(\ded-\frac{\delta^\g\ed + \bed\C}{2}\right) - \frac{1}{2}\left[\bed\left(\B_\mu\B^\mu-\A^2-2\A\C\right) - \bq_\mu\left(4\B_\alpha\C^{\alpha\mu} - 2\C\B^\mu\right) + \bST_{\mu\nu}\left(2\C^{\mu\alpha}\C_\alpha{}^\nu - \C^{\mu\nu}\C\right)\right].
\end{split}
\end{equation}
\end{widetext}

The above derivation is valid for an arbitrary background metric. However, as it is well known, Einstein's equations relate the symmetries of the metric with symmetries of the energy-momentum tensor. In particular, while considering a FLRW metric, the perfect fluid shall have $\bed_{\scp\mu} = \bp_{\scp\mu} = 0$. In addition, through the thermodynamic relations we shall also have $\ben_{\scp\mu} = \bden_{\scp\mu} = 0$, as long as $\bp_{\scp\mu} = \bcs^2\bed_{\scp\mu} + \btal\ben_{\scp\mu}$ and $\bed_{\scp\mu} = \bent\den_{\scp\mu}+ \bden\btemp\ben_{\scp\mu}$.

However, for a barotropic fluid where $\btal = 0$, we no longer have the constraint over the entropy $(\ben_{\scp\mu} =0)$. Notwithstanding, there is still the constraint $\bent\bden_{\scp\mu} + \bden\btemp\ben_{\scp\mu} = 0$.

Thus, from Eq.~\eqref{eq:def:ther12}, we shall have $\btemp = \bent \frac{\dd F(\ben)}{\dd\ben}$, where $F(\ben)$ is an arbitrary function of the entropy. In this case, the constraint amounts to $[\log(\bden) + F(\ben)]_{\scp\mu} = 0$. Therefore, we see that the entropy and the particle density can have nonzero gradient as long as the above combination remains constant.

The covector field $\bent_\mu$ shall be proportional to the normal field. Consequently, we also have that $\bhp{\bent_\mu} = \bfpa_{\scp\mu} + \bfpb\bfpc_{\scp\mu} + \bfpd\ben_{\scp\mu} = 0$. Therefore, the choice of a FLRW as the background metric does not directly imply that all the background scalar fields $\mf_a$'s shall be gradientless.

\end{document}